\documentclass[aps,twocolumn,prx]{revtex4-2}

\usepackage{graphicx,amsmath, amsfonts,bm, color}

\usepackage{xcolor,hyperref}
\hypersetup{
   colorlinks,
   linkcolor={blue!50!black},
   citecolor={blue!50!black},
   urlcolor={blue!80!black}
}
\usepackage{enumitem}

\renewcommand{\=}{\!=\!}
\newcommand{\1}{^{\mbox{\tiny (1)}}}

\newcommand{\tr}{\operatorname{tr}}
\DeclareMathOperator{\sgn}{sgn}

\begin{document}
\title{Velocity-driven frictional sliding: Coarsening and steady-state pulse trains}
\author{Thibault Roch$^{1}$}
\author{Efim A.~Brener$^{2,3}$}
\author{Jean-Fran\c{c}ois Molinari$^{1}$}
\thanks{jean-francois.molinari@epfl.ch}
\author{Eran Bouchbinder$^{4}$}
\thanks{eran.bouchbinder@weizmann.ac.il}
\affiliation{$^{1}$Civil Engineering Institute, Materials Science and Engineering Institute, Ecole Polytechnique F\'ed\'erale de Lausanne, Station 18, CH-1015 Lausanne, Switzerland\\
$^{2}$Peter Gr\"unberg Institut, Forschungszentrum J\"ulich, D-52425 J\"ulich, Germany\\
$^{3}$Institute for Energy and Climate Research, Forschungszentrum J\"ulich, D-52425 J\"ulich, Germany\\
$^{4}$Chemical and Biological Physics Department, Weizmann Institute of Science, Rehovot 7610001, Israel}

\begin{abstract}
Frictional sliding is an intrinsically complex phenomenon, emerging from the interplay between driving forces, elasto-frictional instabilities, interfacial nonlinearity and dissipation, material inertia and bulk geometry. We show that homogeneous rate-and-state dependent frictional systems, driven at a prescribed boundary velocity --- as opposed to a prescribed stress --- in a range where the frictional interface is rate-weakening, generically host self-healing slip pulses, a sliding mode not yet fully understood. Such velocity-driven frictional systems are then shown to exhibit coarsening dynamics saturated at the system length in the sliding direction, independently of the system's height, leading to steadily propagating pulse trains. The latter may be viewed as a propagating phase-separated state, where slip and stick characterize the two phases. While pulse trains' periodicity is coarsening-limited by the system's length, the single pulse width, characteristic slip velocity and propagation speed exhibit rich properties, which are comprehensively understood using theory and extensive numerics. Finally, we show that for sufficiently small system heights, pulse trains are accompanied by periodic elasto-frictional instabilities.
\end{abstract}

\maketitle

\section{Introduction}
\label{sec:Intro}

Frictional systems are composed of two bodies coupled at a contact interface, formed by compressive forces that hold them together. Frictional motion is typically driven by shear forces that are applied far from the frictional interface. The frictional interface and its spatiotemporal dynamics are generically characterized by strong nonlinearity and dissipation, where the interfacial response depends on the local slip velocity $v$ and on the structural state of the interface, carrying memory of its history~\cite{Ruina1983,Rice1983,Marone1998a,Nakatani2001,Baumberger2006,Dieterich2007,Nagata2012,Bhattacharya2014}. Different parts of the frictional interface are coupled through long-range spatiotemporal interactions mediated by the bodies in contact. The latter correspond to the bulk elastodynamics of the bodies, dependent on their elastic response functions --- which in turn depend also on their geometry --- and on material inertia. Consequently, frictional dynamics inherently emerge from the coupled effects of interfacial and bulk physics~\cite{Barras2019,Bar-Sinai2019,Barras2020}.

The interplay between external driving forces, interfacial nonlinearity and dissipation, material inertia and bulk geometry gives rise to very rich spatiotemporal dynamics, which characterize a wide variety of natural and man-made frictional systems, ranging from geological earthquake faults to a multitude of engineering structures and devices~\cite{Ben-Zion2008,Vanossi2013,armstrong1994survey,wojewoda2008hysteretic,massi2007brake,Tonazzi2013}. Understanding, predicting and controlling frictional dynamics remain major scientific and technological challenges. An inseparable aspect of these challenges is that frictional systems host various spatiotemporal instabilities; among these, the most well-characterized instability is associated with rate-weakening friction, i.e.~with physical situations in which the steady frictional resistance is a decreasing function of the slip velocity $v$~\cite{Yamashita1991,Ben-Zion1997,Ben-Zion2001,Ampuero2002,Lapusta2003,Uenishi2003,Rubin2005,Ampuero2008,Kaneko2008,McLaskey2013,Latour2013,Viesca2016,Viesca2016b,Kaneko2016,Kaneko2017,Aldam2017,Gabriel2012}. This interfacial destabilizing process is counteracted by stabilizing bulk elastic interactions, giving rise to a critical elasto-frictional length for the onset of instability (to be accurately defined below). Such instabilities typically result in rapid slip propagation along frictional interfaces, mediated by rupture modes~\cite{Perrin1995,Beeler1996,Cochard1996,Zheng1998,Nielsen2000,Bizzarri2003,Brener2005,Rubin2009,Ben-David2010a,Gabriel2012,Bar-Sinai2013,Svetlizky2014,Putelat2017}.

Spatiotemporal rupture propagation modes can be generally classified into expanding cracklike rupture fronts and compact self-healing slip pulses~\cite{Heaton1990, Perrin1995,Beeler1996,Cochard1996,Zheng1998,Nielsen2000,nielsen2003self,Brener2018,brantut2019stability,Heimisson2019}. In the former, $v$ at an interfacial position behind the propagating mode remains finite as long as propagation persists, while in the latter, $v$ vanishes over a finite time as propagation persists. The conditions for the emergence of self-healing slip pulses and their properties are not yet fully understood, though it is currently accepted that this mode of frictional rupture propagation is prevalent~\cite{Heaton1990,Somerville1999,Lu2007,Shlomai2016,Melgar2017,Heimisson2019}. For example, it is not yet clear how and under what conditions the very same frictional system can feature both cracklike rupture and slip pulses~\cite{Perrin1995,Beeler1996,Cochard1996,Zheng1998,Nielsen2000,Bizzarri2003,Brener2005,Rubin2009,Ben-David2010a,Gabriel2012,Bar-Sinai2013,Svetlizky2014,Putelat2017}. In this paper, we show that there exist generic, and in fact widely used, external driving forces that may lead to the generation of slip pulses. In particular, we show that velocity-driven frictional sliding --- as opposed to stress-driven frictional sliding --- may give rise to propagating periodic slip pulse trains, whose emerging properties are extensively studied below.
\begin{figure*}[ht!]
\includegraphics[width=1\textwidth]{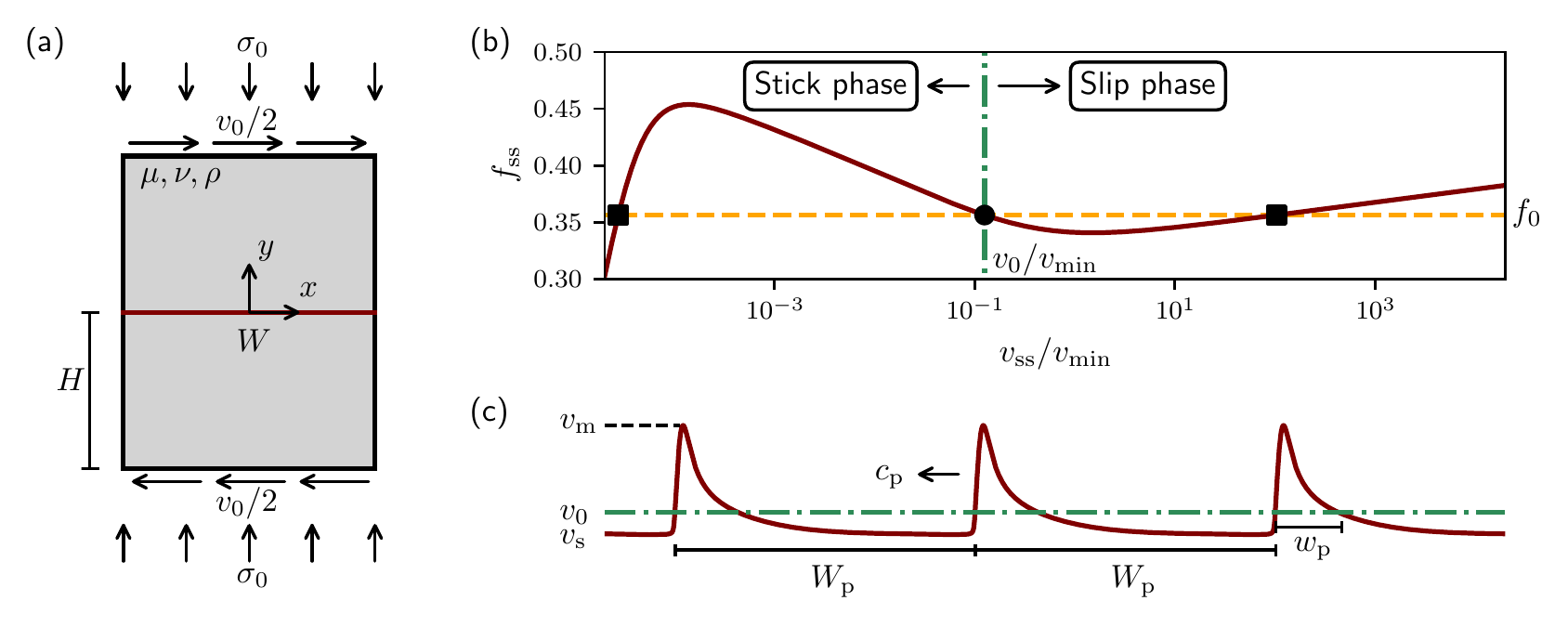}
\caption{(a) A schematic representation of the class of frictional systems under consideration. Two identical elastic bodies of width $W$ and height $H$ are in contact along an interface located at $y\!=\!0$ (brown line). The bodies are under constant normal (compressive) stress $\sigma_0$ and are driven anti-symmetrically at the upper and lower boundaries with a constant shear velocity $v_0/2$ such that the overall applied slip rate is $v_0$. Periodic boundary conditions are imposed in the $x$ direction. (b) The steady-state normalized frictional strength $f_{\rm ss}$ (solid brown line) vs.~the steady-state slip velocity $v_{\rm ss}$ (normalized by $v_{\rm min}$, the minimum of $f_{\rm ss}$) on a semi-logarithmic scale. The curve has a generic N shape~\cite{Bar-Sinai2014}, with two rate-strengthening branches ($df_{\rm ss}/dv_{\rm ss}\!>\!0$) separated by a rate-weakening branch ($df_{\rm ss}/dv_{\rm ss}\!<\!0$), see text for additional discussion. The dashed horizontal orange line represents an imposed driving stress $f_0\!=\!\tau_0/\sigma_0$, which intersects the steady-state friction curve at three points. Two are stable fixed-points on the rate-strengthening branches of the friction law (black squares), while the third one is an unstable fixed-point on the rate-weakening branch (black circle). The dashed-dotted vertical green line corresponds to an imposed driving velocity $v_0$ --- as in panel (a) --- and intersects the friction law at a single point on a rate-weakening branch. As is extensively discussed in the text, imposing a slip velocity $v_0$ on the \emph{unstable} rate-weakening branch cannot result in spatially-homogeneous sliding; rather, it leads to a dynamic and spatially-inhomogeneous phase separation between low velocity (`stick phase', left-pointing arrow) and high velocity (`slip phase', right-pointing arrow) regions. The outcome is a propagating phase-separated state in the form of a pulse train. (c) A generic pulse train observed under velocity-controlled frictional sliding. The train travels at a velocity $c_{\rm p}$, and features periodicity $W_{\rm p}$ and pulses of width $w_{\rm p}$. The pulses feature a maximum slip velocity $v_{\rm m}$ and propagate into the stick phase of a characteristic stick velocity $v_{\rm s}$, which is vanishingly small. The dashed-dotted horizontal green line corresponds to the driving velocity $v_0$.}
\label{fig:fig1}
\end{figure*}

To understand the qualitative differences between velocity-driven and stress-driven frictional sliding, consider the frictional system illustrated in Fig.~\ref{fig:fig1}a, composed of two identical bodies of height $H$ and length $W$, and characterized by elastic constants $\mu$ (shear modulus) and $\nu$ (Poisson's ratio), and mass density $\rho$. The bodies are held together by a homogeneous normal (compressive) stress $\sigma_0$ and are subjected to some shear-related boundary conditions. In the figure, the boundary conditions at the upper and lower edges are denoted by $v_0/2$, implying that in this case the applied velocity (of overall magnitude $v_0$) is kept fixed. However, one could also consider a situation in which the applied shear stress $\tau_0$ is kept fixed.

To explain why these two types of driving forces may lead to qualitatively different physical consequences, we plot in Fig.~\ref{fig:fig1}b the steady-state frictional resistance, $f_{\rm ss}\=\tau_{\rm ss}/\sigma_0$ (where $\tau_{\rm ss}$ is the steady-state frictional strength/stress), as a function of the logarithm of the steady-state slip velocity $v_{\rm ss}$. By steady-state we mean that we focus on a point along the frictional interface that experiences a slip velocity $v_{\rm ss}$ for a sufficiently long time, and measure the resulting frictional strength/stress $\tau_{\rm ss}$; the steady-state friction coefficient, which out of steady-state depends on the instantaneous slip velocity $v$ and on a set of structural (internal) variables (see below), is simply given by $f_{\rm ss}\=\tau_{\rm ss}/\sigma_0$, as stated above. The steady-state friction curve $f_{\rm ss}(v_{\rm ss})$ features a generic N shape~\cite{Bar-Sinai2014}, where friction is rate-strengthening at both low and high slip velocities (the latter occurs above the minimum of the curve, denoted by $v_{\rm min}$), and rate-weakening at intermediate velocities (typically spanning a few orders of magnitude, note again the logarithmic $v_{\rm ss}$-axis).

To highlight the differences between velocity-driven and stress-driven sliding, consider the frictional system illustrated in Fig.~\ref{fig:fig1}a under the application of a fixed (total) slip velocity $v_0$, which resides in the rate-weakening branch of the steady-state friction curve (marked by the dashed-dotted vertical green line in Fig.~\ref{fig:fig1}b). Consider also the stress-driven counterpart of this system, i.e.~the case in which a fixed shear stress $\tau_0$ is applied. The latter corresponds to $f_0\=\tau_0/\sigma_0$ (marked by the dashed horizontal orange line in Fig.~\ref{fig:fig1}b) and is chosen such that $f_0\=f_{\rm ss}(v_0)$, i.e.~the dashed-dotted vertical and dashed horizontal lines intersect the steady-state friction curve at the very same point, marked by a black circle in Fig.~\ref{fig:fig1}b. The horizontal line, corresponding to stress-driven sliding, intersects the steady-state friction curve also at two other points on the two rate-strengthening branches, marked by black squares. Let us focus first on the stress-driven case and ask whether spatially homogeneous and stable steady sliding can emerge under these conditions. It is clear that steady sliding of the whole system at $v_0$ is not possible because the rate-weakening branch is unstable (for $W$ larger than the critical elasto-frictional length). On the other hand, homogeneous and stable steady sliding at the velocities corresponding to the black squares is possible, because the latter reside on rate-strengthening branches, which are stable~\cite{Barras2019}.

This situation is in sharp and qualitative contrast to the velocity-driven case, where the sliding velocity $v_0$ is enforced on the outer boundaries. In this case, spatially homogeneous sliding is \emph{impossible} because $v_0$ resides on the rate-weakening branch, implying instability, and no other spatially homogeneous velocity solutions are possible either (unlike the stress-driven case). Consequently, either steady sliding does not exist at all or spatially inhomogeneous steady-state $v(x,t)$ emerges ($x$ here denotes the spatial coordinate along the frictional interface and $t$ is the time) such that $W^{-1}\!\int_0^W\! v(x,t)\,dx\=v_0$ at any time. In the latter situation, $v(x,t)$ must take the form of a steadily propagating pulse train, illustrated in Fig.~\ref{fig:fig1}c. The pulse train is characterized by a spatial periodicity $W_{\rm p}$ and a propagation velocity $c_{\rm p}$. Each pulse within the train features a characteristic width $w_{\rm p}$ (to be accurately defined below), a maximal velocity $v_{\rm m}\!>\!v_0$ and a minimal velocity $v_{\rm s}\!<\!v_0$. Since $v_{\rm s}$ is typically much smaller than $v_0$, it is termed the `stick velocity', corresponding to a nearly non-sliding state, which is termed the `stick phase' (cf.~Fig.~\ref{fig:fig1}b). The parts of a pulse train that feature $v(x,t)\!>\!v_0$ can be regarded as `slip phases' (cf.~Fig.~\ref{fig:fig1}b); consequently, a pulse train may be viewed as a propagating mode composed of alternating stick and slip phases.

This physical picture of pulse trains as propagating modes composed of alternating stick and slip phases, emerging under velocity-driven conditions, may suggest an analogy to phase separation (and the associated Maxwell construction) in equilibrium thermodynamics~\cite{baus2007equilibrium}. In the latter, phase separation emerges when the pressure-volume isotherm of a system in equilibrium features a non-monotonic behavior, corresponding to a non-convex free energy~\cite{baus2007equilibrium}. If then the system is enforced to have a volume in the non-monotonic region of the pressure-volume isotherm, where the pressure is an increasing function of the volume (thermodynamic stability requires the pressure to be a decreasing function of the volume), it cannot attain a (single) homogeneous phase; rather, the system undergoes a phase transition that leads to the co-existence of two phases of different densities~\cite{baus2007equilibrium}. While $f_{\rm ss}(v_{\rm ss})$ of Fig.~\ref{fig:fig1}b is by no means an equilibrium pressure-volume isotherm, rather it corresponds to a strongly dissipative interfacial response function of a driven open system, and while the imposed slip velocity $v_0$ inside the rate-weakening (unstable) branch is by no means the volume of an equilibrium system, there exists a clear and direct analogy between the two physical situations. In the frictional case, the result is not static phase separation, but rather a periodic and propagating phase-separated state of alternating stick and slip phases, the train of pulses illustrated in Fig.~\ref{fig:fig1}c.

It is important to stress that in practical terms velocity-driven frictional sliding is the rule, rather than the exception. That is, velocity is almost always what is actually being controlled; in order to maintain a fixed stress, one needs to employ a feedback loop such that the velocity is precisely varied so as to keep the stress fixed. This procedure depends on the dynamics of the system, which may be fast, and is in general difficult to achieve. In most cases, it is just the velocity which is prescribed.

Our goal in this paper is to understand the spatiotemporal dynamics of velocity-driven frictional systems and in particular the emergence of steadily propagating pulse trains. Moreover, we aim at understanding the selection of the train properties, i.e.~its spatial periodicity and propagation velocity, as well as the properties of a single pulse within the train. These goals are achieved using extensive numerical simulations --- employing several computational methodologies --- and theoretical analysis performed within a generic rate-and-state friction constitutive framework (to be detailed below). In particular, we employ the Boundary Integral Method in the $H\!\to\!\infty$ limit; while this method is formulated for stress-driven sliding in terms of $\tau_0$~\cite{Geubelle1995,Morrissey1997,Breitenfeld1998}, we show that it can nevertheless be employed to mimic velocity-driven frictional dynamics. For finite $H$, we employ the Finite Element Method, as explained below.

We find that frictional systems under velocity-driven conditions, where the imposed velocity $v_0$ resides on an unstable rate-weakening branch of the steady-state friction curve, feature coarsening dynamics that lead to pulse trains whose periodicity $W_{\rm p}$ is determined by the system length $W$ (where periodic boundary conditions in the sliding direction are employed, see below), independently of $H$. We also show that in the small $H$ limit, coarsening competes with elasto-frictional instabilities, giving rise to pulse trains with $W_{\rm p}\=W$ that experience repeated/periodic instabilities. Furthermore, we show that the pulse train propagation velocity $c_{\rm p}$, the average single pulse slip velocity and the single pulse width $w_{\rm p}$ are related through an equation of motion inspired by fracture mechanics~\cite{Freund1998}. The latter involves the pulse leading edge singularity and the emergence of an effective fracture energy. Finally, we show that the single pulse width $w_{\rm p}$ reveals non-trivial dependencies on $W_{\rm p}\=W$ and $v_0$, featuring properties that are qualitatively different from those of ideal pulse solutions~\cite{Kostrov1964,freund1979mechanics,Broberg1999}. Taken together, we provide a comprehensive physical picture of velocity-driven frictional sliding in general, and of the properties of the emerging pulse trains in particular.

\section{Simulating velocity-driven frictional dynamics}
\label{sec:simulations}

The rich spatiotemporal dynamics featured by frictional systems, with the multitude of physical factors at play (as discussed above), make purely analytical treatments of this class of problems practically impossible. Consequently, one needs to resort to numerical simulations, at least at the initial stages of investigating a given set of questions. Independently of the approach taken --- either analytic, computational or hybrid --- one should first adopt bulk and interfacial constitutive relations, which is done in Sect.~\ref{subsec:RSF}. To fully define the problem at hand, one should then specify the bulk geometry and external driving forces (boundary conditions) --- here following Fig.~\ref{fig:fig1}a ---, and finally one needs to select a solution method for the coupled bulk-interface problem. Focusing first on computational methods, we explain in Sect.~\ref{subsec:BIM} how the Boundary Integral Method --- conventionally formulated in terms of stress boundary conditions --- can be used to mimic velocity-driven frictional dynamics in the $H\!\to\!\infty$ limit. Next, in Sect.~\ref{subsec:FEM}, we discuss the usage of the Finite Element Method to address the finite $H$ regime.

\subsection{Bulk and interfacial constitutive relations: Linear elastodynamics and rate-and-state friction}
\label{subsec:RSF}

The frictional system illustrated in Fig.~\ref{fig:fig1}a is formed by two symmetric bodies, each satisfying its own continuum momentum balance equation $\rho\,\ddot{\bm u}(x,y,t)\=\nabla\!\cdot{\bm \sigma}(x,y,t)$, where $\rho$ is the mass density, $\bm u(x,y,t)$ is the displacement vector field, $\bm \sigma(x,y,t)$ is the stress tensor field, $(x,y)$ is a two-dimensional Cartesian coordinate system and $t$ is the time (a superposed dot represents a partial time derivative). $\bm \sigma$ in each body is related to $\bm u$ through a bulk constitutive relation, which is taken here to be that of linear elasticity, i.e.~we adopt Hooke's law~\cite{Landau1986} of the form $(1+\nu)\mu\left[\nabla {\bm u}\!+\! (\nabla {\bm u})^{\mbox{\scriptsize T}}\right]\!=\!{\bm \sigma}-\nu({\bm I}\tr{\bm \sigma}-{\bm \sigma})$. Here ${\bm I}$ is the identity tensor, $\nu$ is Poisson's ratio and $\mu$ is the shear modulus of each body. Note that body forces are neglected in the momentum balance equation and that the interface resides at $y\=0$ (cf.~Fig.~\ref{fig:fig1}a).
An interfacial constitutive law is an implicit boundary condition for the two bulk problems defined above, formulated in terms of a functional relation between the interfacial shear stress, the interfacial normal stress and the slip velocity (and typically also additional interfacial state fields, see below). In the problems considered in this paper, the interfacial normal stress is constant (uncoupled to frictional sliding), i.e.~$\sigma_{yy}(x,y\=0,t)\=-\sigma_0$, where $\sigma_0$ is the applied compressive stress (cf.~Fig.~\ref{fig:fig1}a).

The slip velocity $v(x,t)$ is the time derivative of the slip displacement $\delta(x,t)$, $v(x,t)\=\dot\delta(x,t)$. Under in-plane shear (the so-called mode-II) conditions, where ${\bm u}(x,y,t)\=(u_x(x,y,t), u_y(x,y,t),0)$ (here the $z$ component of the displacement vector field, $u_z(x,y,t)$, vanishes), one has $\delta(x,t)\!\equiv\!{u}_x(x,y\!\to\!0^+,t)-{u}_x(x,y\!\to\!0^-,t)$, where $+/-$ correspond to the upper/lower bodies, respectively. The relevant interfacial shear stress in this case is $\sigma_{xy}(x,y\=0,t)$. Under anti-plane shear (the so-called mode-III) conditions, where ${\bm u}(x,y,t)\=(0,0,u_z(x,y,t))$ and $z$ is the out-of-plane direction (perpendicular to the $x\!-\!y$ plane), one has $\delta(x,t)\!\equiv\!{u}_z(x,y\!\to\!0^+,t)-{u}_z(x,y\!\to\!0^-,t)$, and relevant interfacial shear stress in this case is $\sigma_{yz}(x,y\=0,t)$. The interfacial shear stress, either $\sigma_{xy}(x,y\=0,t)$ (mode-II) or $\sigma_{yz}(x,y\=0,t)$ (mode-III), is continuous across the interface and equals the frictional stress/strength $\tau(x,t)$. Below we present results for both mode-II and mode-III, which are qualitatively and even semi-quantitatively similar. Mode-III (corresponding to scalar elastodynamics) is, however, mathematically and computationally simpler.

The frictional stress/strength $\tau(x,t)$ is related to $v(x,t)$, $\sigma_{yy}(x,y\=0,t)$ (which in our case equals $-\sigma_0$) and additional interfacial state fields through the interfacial constitutive law. The latter, at any position $x$ along the interface and at any time $t$, is described by the following local relation
\begin{equation}
\label{eq:friction_law}
\tau(\sigma_0,v,\phi)=\sigma_0\,\sgn(v)\,f(|v|,\phi) \ ,
\end{equation}
where $\phi(x,t)$ is a non-equilibrium order parameter, sometimes termed an internal-state field, which represents the structural state of the interface and encodes its history~\cite{Ruina1983,Rice1983,Marone1998a,Nakatani2001,Baumberger2006,Dieterich2007,Nagata2012,Bhattacharya2014}. Extensive evidence indicates that $\phi$ physically represents contact's age/maturity~\cite{Rice1983,Marone1998a,Nakatani2001,Baumberger2006,Dieterich2007,Nagata2012,Bhattacharya2014} and that its evolution follows
\begin{equation}
\label{eq:dot_phi}
\dot\phi = g\!\left(\frac{\phi|v|}{D}\right) \ ,
\end{equation}
with $g(1)\=0$.
The characteristic slip displacement $D$ controls the transition from a stick state $v\!\approx\!0$ to a steadily slipping/sliding state $v_{\rm ss}\!\ne\!0$, with $\phi_{\rm ss}\=D/v_{\rm ss}$ (the latter corresponds to $\dot\phi\=g(1)\=0$). Under steady-state sliding conditions and a controlled normal stress $\sigma_0$, the function $f_{\rm ss}(v_{\rm ss})\=f(|v_{\rm ss}|,\phi_{\rm ss}\=D/v_{\rm ss})\=\tau_{\rm ss}(v_{\rm ss})/\sigma_0$ has been measured over a broad range of slip rates $v$ for many materials~\cite{Bar-Sinai2014}.
$f_{\rm ss}(v_{\rm ss})$ is generically N-shaped, as shown in Fig.~\ref{fig:fig1}b, where the precise functional form of $f(\cdot)$ and $g(\cdot)$ are detailed in Appendix~\ref{app:friction_laws}.

\subsection{Mimicking velocity-driven frictional dynamics in infinite systems using the boundary integral method}
\label{subsec:BIM}

With the bulk and interfacial constitutive relations at hand, the coupled bulk-interface problem is fully defined once the bulk geometry is specified. The system length is taken to be $W$, as shown in Fig.~\ref{fig:fig1}a; in order to avoid lateral edge effects, we employ periodic boundary conditions along this direction. The remaining geometric length scale in the problem is $H$, i.e.~the height of each body (cf.~Fig.~\ref{fig:fig1}a). It makes a difference whether $H$ is taken to be arbitrarily large or finite, as is explained below. Here we first discuss the $H\!\to\!\infty$ case.

In the $H\!\to\!\infty$ limit, i.e.~when the upper and lower boundaries are located infinitely away from the frictional interface, and no wave reflections from these boundaries take place, on can naturally invoke a Green's function approach. The latter allows to eliminate the two bulk problems altogether, reducing the coupled bulk-interface problem to an interfacial integro-differential equation of the form~\cite{Geubelle1995,Ben-Zion1997,Morrissey1997,Breitenfeld1998,Lapusta2000}
\begin{equation}
\label{eq:BIM}
\tau[v(x,t),\phi(x,t)]=\tau_0(t) - \frac{\mu}{2c_s}[v(x,t)-v_0]+s(x,t) \ .
\end{equation}
where $\phi(x,t)$ satisfies Eq.~\eqref{eq:dot_phi} and the right hand side is the interfacial shear stress. The latter contains three physically distinct contributions, to be discussed next.

The first contribution is the applied (spatially homogeneous) stress $\tau_0(t)$, i.e.~this approach is directly applicable to stress-controlled conditions. The second contribution is the so-called radiation damping term~\cite{Ben-Zion1995,Perrin1995,Zheng1998,Crupi2013}, where $c_s$ is the shear wave speed and $v_0$ is set as a reference slip velocity (to be identified with the applied slip velocity $v_0$ in our velocity-controlled setting, see below). The radiation damping term locally depends on $v(x,t)$ and physically represents plane-waves being radiated away from the interface into the surrounding bulks, serving as effective damping from the perspective of the interface. Finally, the third contribution $s(x,t)$ is non-local in space and time, and physically represents the spatiotemporal interaction of different points on the interface, mediated by bulk elastodynamics. $s(x,t)$ generally does not admit real-space representation, and is related to the gradient of $\delta(x,t)$ in the spectral domain via a convolution integral, which is different for mode-II and mode-III~\cite{Geubelle1995,Morrissey1997,Breitenfeld1998}. The spectral nature of the formulation fits the choice of periodic boundary condition in the lateral/sliding direction (with periodicity $W$), and is reflected in its common name, the spectral Boundary Integral Method (BIM).

As explained above, and as evident in Eq.~\eqref{eq:BIM}, the BIM is most suitable for stress-controlled boundary conditions represented by $\tau_0(t)$. Yet, we propose here an approach in which the BIM formulation can be nevertheless used to mimic velocity-driven frictional dynamics that are of interest here. The idea is the following: as explained above, under velocity-driven conditions \emph{and} once the system reached steady-state, one has
\begin{equation}
\label{eq:avr_v}
\frac{1}{W}\int_0^W\! v(x,t)\, dx = v_0 \ .
\end{equation}
Consequently, one can choose $\tau_0(t)$ --- an a priori unknown function of time $t$ --- such that Eq.~\eqref{eq:avr_v} is satisfied at any time $t$. That is, we propose to treat $\tau_0(t)$ in Eq.~\eqref{eq:BIM} as unknown and instead to impose Eq.~\eqref{eq:avr_v} for any $t$. Thinking about a numerical implementation of the formulated problem, where both the time $t$ and the spatial coordinate $x$ are discretized, it is clear that the above suggestion leads to a well-defined problem; at each discrete time $t_i$, we added a single unknown $\tau_0(t_i)$ and a single constraint (Eq.~\eqref{eq:avr_v} at $t_i$). Moreover, as $v_0$ is the relevant slip velocity in this modified BIM formulation, we used it as a reference velocity in the radiation damping term in Eq.~\eqref{eq:BIM}.

The modified BIM formulation, aimed at mimicking velocity-driven frictional sliding, has both clear advantages and potential limitations. On the one hand, it is a relatively computationally cheap and very robust approach, which is expected to reveal velocity-driven steady-states in the $H\!\to\!\infty$ limit (if these exist). On the other hand, as Eq.~\eqref{eq:avr_v} is strictly valid only in steady-state --- i.e.~some deviations from it are expected in early-time, out of steady-state dynamics --- some dynamical aspects of the full velocity-driven problem may not be accurately captured. We address this potential limitation in two ways; first, we mainly focus on the long-time, steady-state behavior of the system, where our modified BIM formulation is strictly valid. Second, we verify through Finite Element Method (FEM) that the possible deviations of the transient dynamics in the $H\!\to\!\infty$ BIM calculations from the exact transient dynamics have no effect on the obtained steady-state solutions. More importantly, the FEM formulation to be discussed next allows us to understand the roles played by a finite $H$ on the physics of the problem at hand.

\subsection{Simulating velocity-driven frictional dynamics in finite systems using the finite element method}
\label{subsec:FEM}

The length scale $H$ may play important roles in velocity-driven frictional sliding. Consequently, it is essential to supplement the $H\!\to\!\infty$ BIM calculations with FEM ones, which allow to probe the finite $H$ regime. Moreover, finite $H$ FEM calculations can be used to verify the validity of the modified BIM calculations, as discussed above, for selected test cases in which $H$ is chosen to be large enough (see below). The new physics introduced by a finite $H$ is the wave interaction of the frictional interface with the boundaries at $y\=\pm H$, which is absent in the $H\!\to\!\infty$ limit, where wave are being radiated from the interface to $\pm\infty$ without reflections.

FEM simulations of frictional systems, especially when rate-and-state friction is taken into account, are significantly more challenging and computationally demanding than their BIM counterparts~\cite{Rezakhani2020}. They are also prone to numerical and physical instabilities (associated with the finite $H$) that are absent in their BIM counterparts. One such physical instability is encountered in our large-$H$ FEM calculations. This issue, along with the full details of our FEM simulations, are addressed in Appendix~\ref{app:FEM}. Agreement between our FEM and BIM calculations for mode-II (simpler mode-III calculations are performed only using BIM) is demonstrated in the next section, where we start discussing the emergent physics of velocity-driven frictional systems, and pulse trains in particular.

\section{Coarsening dynamics: The selection of pulse train periodicity}
\label{sec:coarsening}

The first question we aim at addressing is whether steady-state pulse trains indeed emerge under velocity-driven sliding conditions and if so, what determines the train periodicity $W_{\rm p}$. As the latter is of length dimension, one can a priori ask what quantities of length dimension existing in the posed problem could possibly determine $W_{\rm p}$. The problem at hand, as formulated above, features 3 length scales: the system height $H$, the system length $W$ and in addition to these two geometric/extrinsic length scales, we also have the intrinsic interfacial length $D$. As $D$ is a mesoscopic length that characterizes the interfacial response, it does not appear in itself in the macroscopic coupled bulk-interface problem. Rather, it appears through the elasto-frictional length $L_{\rm c}$ that characterizes the rate-weakening ($df_{\rm ss}/dv\!<\!0$) elasto-frictional instability~\cite{Ruina1983,Yamashita1991,Ben-Zion1997,Ben-Zion2001,Ampuero2002,Lapusta2003,Uenishi2003,Rubin2005,Ampuero2008,Kaneko2008,McLaskey2013,Latour2013,Viesca2016,Viesca2016b,Kaneko2016,Kaneko2017,Aldam2017,Gabriel2012}.

The elasto-frictional length $L_{\rm c}$ can be determined using a systematic linear stability analysis~\cite{Aldam2017,Bar-Sinai2019}, which is very briefly reviewed in Appendix~\ref{app:nucleation_length}. Here we are mainly interested in the scaling structure of $L_{\rm c}$, taking the form
\begin{equation}
\label{eq:Lc}
L_{\rm c}(H,D) = H \,{\cal F}\!\left(\!\frac{\mu\, D}{-\sigma_0 H v_0\,df_{\rm ss}/dv_0}\!\right) \ ,
\end{equation}
where ${\cal F}({\cal X})\!\sim\!{\cal X}$ for ${\cal X}\!\ll\!1$ and ${\cal F}({\cal X})\!\sim\!\sqrt{\cal X}$ for ${\cal X}\!\gg\!1$~\cite{Bar-Sinai2019}, with ${\cal X}\!\equiv\!-\mu D/(\sigma_0 H v_0 \,df_{\rm ss}/dv_0)$. ${\cal X}$ clearly manifests the coupled bulk-interface nature of $L_{\rm c}$, incorporating the shear modulus $\mu$ of the bodies in contact, their height $H$ (a geometric/extrinsic length), the applied normal stress $\sigma_0$, the mesoscopic interfacial length $D$ and the constitutive interfacial property $v_0\,df_{\rm ss}/dv_0\=\tfrac{df_{\rm ss}(v_0)}{d\!\log{v}}\!<\!0$. Note that the limiting behaviors of ${\cal F}({\cal X})$, stated above, imply that $L_{\rm c}$ is independent of $H$ in the limit $H\!\to\!\infty$. The length $L_{\rm c}$ implies that infinitesimal perturbations of wavelength larger than $L_{\rm c}$, on top of homogeneous sliding with slip velocity $v_0$, are linearly unstable. That is, for $L_{\rm c}\!<\!W$ the system is linearly unstable as unstable perturbations can fit in (to ensure that this is the case, we use $L_{\rm c}\!\ll\!W$ throughout this work). Moreover, the wavelength of the fastest growing mode of instability features the same scaling properties as $L_{\rm c}$ in Eq.~\eqref{eq:Lc}.

\begin{figure}[ht!]
\includegraphics[width=0.5\textwidth]{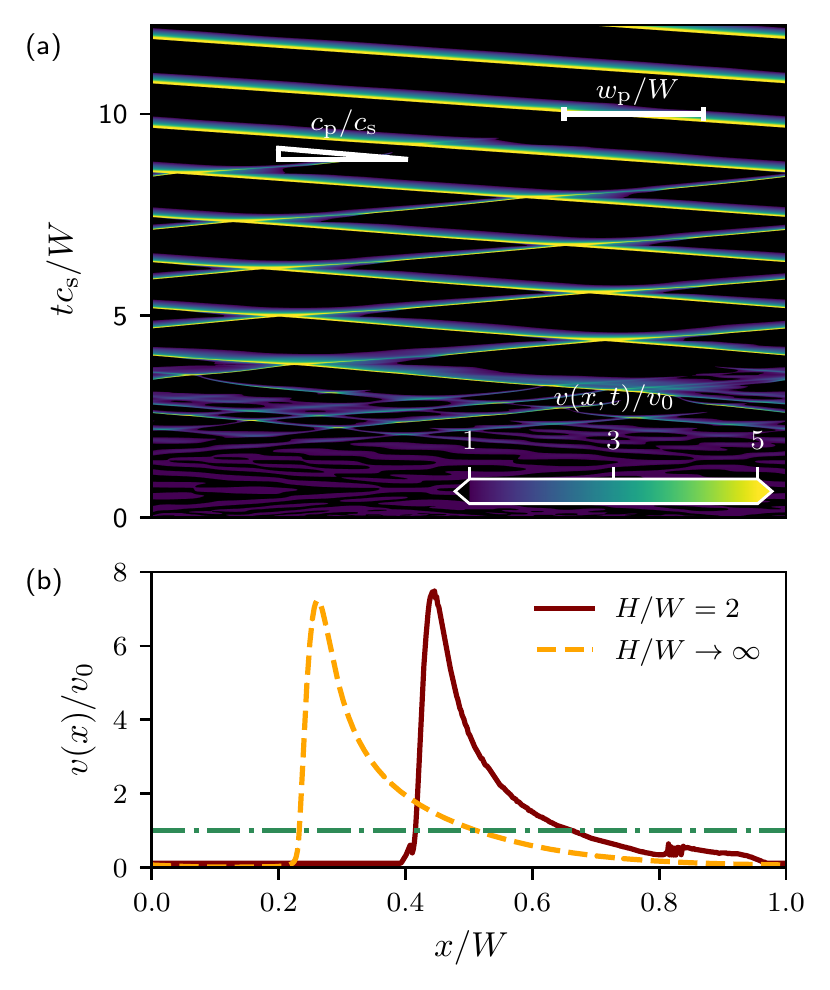}
\caption{(a) A space-time plot of the slip velocity field $v(x,t)$ (normalized by $v_0$) obtained using a BIM simulation ($H\!\to\!\infty$) with $v_0\!=\!3 \!\times\! 10^{-3} \text{m/s} $, $W\!=\!28.14\text{m}$ and the N-shaped steady-state friction law of Fig.~\ref{fig:fig1}b. Black regions correspond to the stick phase, $v(x,t)\!<\!v_0$. At early times, the system host several instabilities. The interface progressively coarsens with increasing time until a single pulse remains in the periodic domain. The asymptotic pulse propagation velocity $c_{\rm p}/c_{\rm s}$ is shown as the slope of the slip phase $v(x,t)\!>\!v_0$. The normalized pulse width $w_p/W$ corresponds to the extent of the slip phase along the $x$ axis at a given time $t$. A movie of this simulation is available in~\cite{Movies}. (b) A snapshot of $v(x,t)$ of a steady-state pulse train, comparing a BIM simulation (dashed orange line) and a FEM simulation (solid brown line) with $v_0\!= \!1\!\times\! 10^{-3} \text{m/s}$, $W\!=\!6\text{m}$ and the steady-state friction law of Fig.~\ref{fig:figSM1}. The dashed-dotted horizontal green line corresponds to the driving velocity $v_0$. Both pulses are propagating towards the left at the same velocity $c_{\rm p}\!\simeq\!0.35 c_s$ and their shapes are almost identical.}
\label{fig:fig2}
\end{figure}

To address the question of whether $W_{\rm p}$ exists, i.e~whether steady-state pulse trains emerge, and if so how $W_{\rm p}$ depends on $L_{\rm c}(H,D)$, $H$ and $W$, we first explore the $H\!\to\!\infty$ limit using the BIM formulation of Sect.~\ref{subsec:BIM}. That is, we first consider the case $L_{\rm c}\!\ll\!W\!\ll\!H$. A representative result under mode-III conditions is presented in Fig.~\ref{fig:fig2}a, where a space-time plot of $v(x,t)/v_0$ is shown. It is observed that, as expected, at early times the velocity-driven frictional system experiences instabilities, resulting in multiple interacting slip pulses. As time progresses (increasing vertical direction in the space-time plot), the number of pulses decreases until --- in the long-time limit --- a single pulse survives. The latter corresponds to a steady-state pulse train of periodicity $W$, i.e.~$W_{\rm p}\=W$ (recall that throughout this work we employ periodic boundary conditions along the sliding direction). That is, the frictional system undergoes coarsening dynamics  that are saturated  at  the  system length $W$. To test the robustness of this observation, we varied $W$ and the interfacial parameters over a wide range (cf.~Appendix~\ref{app:friction_laws} and~\ref{app:BIM}); independently of these variations, we always observed $W_{\rm p}\=W$ in our $H\!\to\!\infty$ calculations, indicating coarsening dynamics that are truncated at the system length.

Next, we aim at understanding whether a finite height $H$ can result in a qualitative change in this physical picture, i.e.~whether the pulse train periodicity $W_{\rm p}$ can be affected/determined by $H$. To address this question, one clearly needs to resort to FEM calculations. Due to numerical stability considerations, we use a rate-and-state friction law that does not feature the very low velocity rate-strengthening branch, see Appendix~\ref{app:ws_friction} for discussion and details. In particular, we focus on the regime where $H\!\simeq\!W\!\gg\!L_{\rm c}$ in order to see whether $H$ affects $W_{\rm p}$ in addition to $W$ (as they are comparable). In this regime, we encountered in addition to the elasto-frictional instability associated with $L_{\rm c}$ also a system-size instability. In this instability, the slip velocity vanishes throughout the interface for some period of time, and then interface starts sliding almost \emph{homogeneously}. That is, this instability appears to feature a space-independent stick-slip behavior (see Appendix~\ref{app:mode_0}), which is likely to be related to the finite-$H$ elasto-frictional instabilities discussed in~\cite{Brener2016}. We have not studied this instability in depth as we strongly suspect that it does not manifest itself in the limit of large $W$ (which is not accessible in our FEM calculations) and large times, i.e.~it does not affect the long-time behavior of the system. Instead, we effectively eliminated this space-independent instability by breaking translational symmetry along the interface using a constraint similar to Eq.~\eqref{eq:avr_v}, as explained in Appendix~\ref{app:mode_0}.
\begin{figure*}[ht!]
\includegraphics[width=1\textwidth]{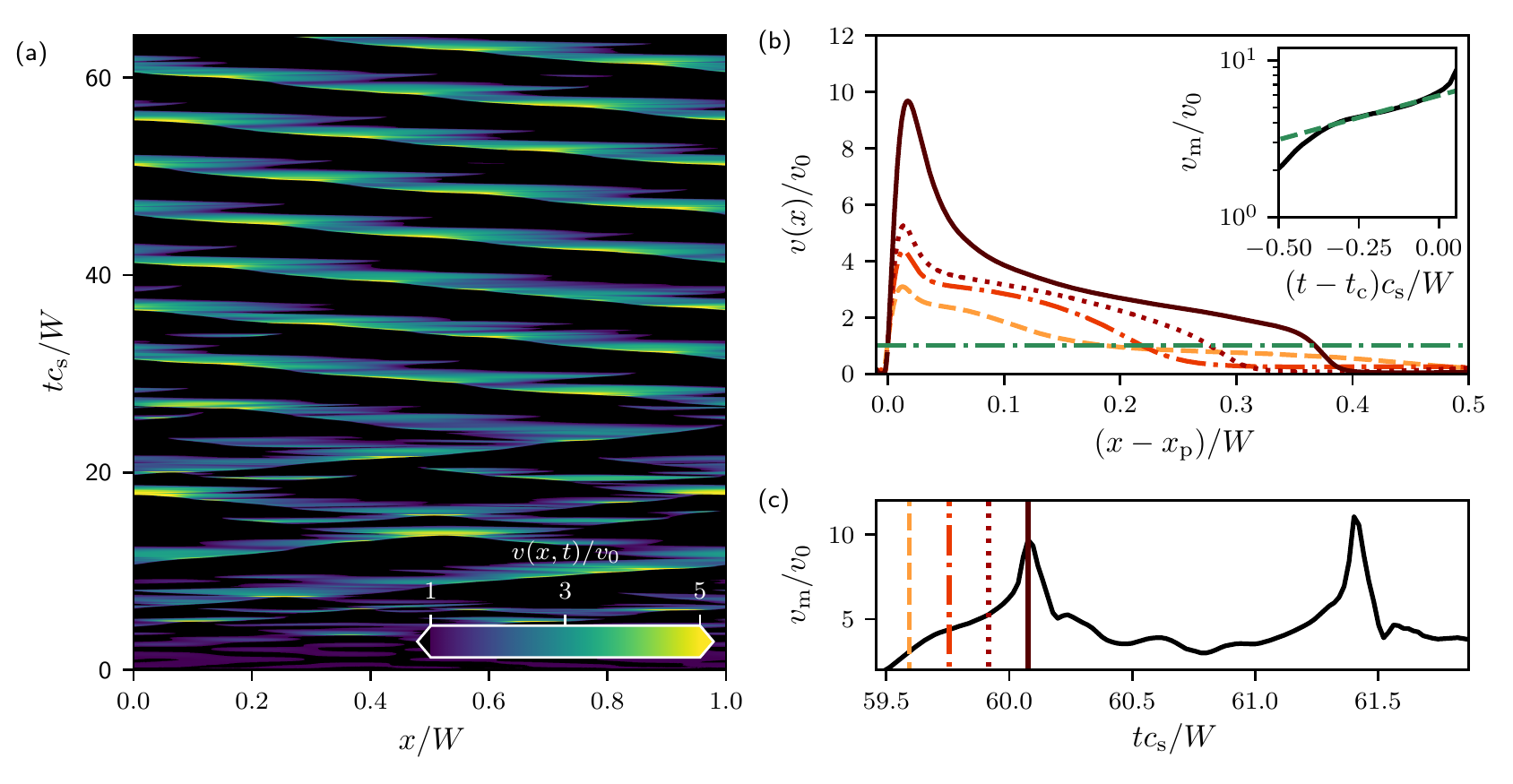}
\caption{(a) A space-time plot as in Fig.~\ref{fig:fig2}a for a FEM simulation with $v_0\!=\!2\!\times\!10^{-3} \text{m/s} $, $W\!=\!4\text{m}$, $H\!=\!0.2 \text{m}$ and the dashed-dotted orange steady-state friction law of Fig.~\ref{fig:figSM1}. The observed behavior is similar to the BIM example ($H\!\to\!\infty$) of Fig.~\ref{fig:fig2}a, with the notable difference that the pulse train that remains in the system in the long-time coarsening limit features repeated oscillations (visible from the variation of the width and slip rate of the pulse). A movie of this simulation is available in~\cite{Movies}. (b) Successive snapshots of $v(x)/v_0$ are presented in the co-moving frame of the pulse train (with $x_{\rm p}$ being the pulse's leading edge). The pulse profiles are ordered in time as follows (line colors): dashed yellow, dashed-dotted orange, dotted brown, and solid dark brown. The dashed-dotted horizontal green line corresponds to the driving velocity $v_0$. As time progresses, both the pulse width $w_{\rm p}$ and the slip velocity increase, indicating an elasto-frictional instability. A movie of the instability in the co-moving frame of the pulse is available in~\cite{Movies}. (inset) The time evolution of the normalized maximum velocity $v_{\rm m}$ (solid black curve) of the dynamics shown in panel (b), presented on a semi-logarithmic scale. The green dashed line is the best linear fit, corresponding to an exponential growth characterizing a linear instability, with the growth rate being the slope of this line. $t_{\rm c}$ corresponds to the time at which the growth becomes nonlinear. (c) The time evolution of $v_{\rm m}/v_0$. The vertical lines indicate the time of the snapshots in panel (b), with the same style and color codes. Two events of this repeated behavior (growth and decay of the pulse) are shown.}
\label{fig:fig3}
\end{figure*}

In Fig.~\ref{fig:fig2}b, we present a snapshot of the long-time behavior of an FEM simulation (as described above) of a frictional system with $H\=2W$ (solid line), where yet again a steady-state pulse train with $W_{\rm p}\=W$ is observed. This result is further strengthened by additional FEM calculations where the value of $H$ has been varied for a fixed $W$, still having $H\!\simeq\!W\!\gg\!L_{\rm c}$, yielding $W_{\rm p}\=W$ independently of $H$. The independence of $W_{\rm p}$ on $H$ also suggests that for a fixed set of frictional parameters and $W$, the $H\=2W$ FEM calculation and the $H\!\to\!\infty$ BIM calculation should give rise to very similar results. Such a quantitative agreement is possible if the radiation damping term is not too large. This is the case if $c_{\rm p}$ is not close to the relevant wave speed.

In Fig.~\ref{fig:fig2}b, we superimposed the $H\!\to\!\infty$ BIM results (dashed line, for mode-II as in the FEM calculation) on top of the $H\=2W$ FEM ones (the two snapshots are shifted along $x$ for visual clarity). It is observed that in both cases $W_{\rm p}\=W$ and that the pulse shape is almost identical (the same applies to the pulse train velocity $c_{\rm p}\!\simeq\!0.35c_s$, see figure caption). These convergent results demonstrate the robustness and validity of our FEM calculations. More importantly, they show that velocity-driven frictional systems exhibit coarsening dynamics that are saturated at the system length, independently of the system’s height $H$ in the regime $L_{\rm c}\!\ll\!W\!\le\!H$, leading to steadily propagating pulse trains with $W_{\rm p}\=W$. In light of the independence of $W_{\rm p}$ on $H$, we would like next to understand whether $L_{\rm c}$ can affect $W_{\rm p}$, which requires to consider the small $H$ regime.

\section{The small $H$ limit: The competition between coarsening and elasto-frictional instabilities}
\label{sec:smallH}

The results of the previous section strongly suggest that we have $W_{\rm p}\=W$ independently of $H$ for $L_{\rm c}\!\ll\!W\!\le\!H$, where $L_{\rm c}$ does not seem to play a role in the long-time pulse train as well. To address the possible role of $L_{\rm c}$ in the selection of the pulse train periodicity $W_{\rm p}$, we explore here the small $H$ limit for which $H\!<\!L_{\rm c}\!\ll\!W\!$. Before discussing our FEM calculations in this regime, let us invoke some theoretical considerations regarding their possible outcomes. One possibility is that we find steady-state pulse trains with $W_{\rm p}\!\simeq\!L_{\rm c}$. While we cannot a priori rule out this possibility, we note that --- if true --- it implies that coarsening dynamics play no role whatsoever in this regime. It is not easy to imagine this in light of the fact that coarsening dynamics strongly dominate the physics for $L_{\rm c}\!\ll\!W\!\le\!H$. Another possibility is that coarsening dynamics remain dominant, leading to $W_{\rm p}\=W$ as before. This, however, cannot lead to a strict steady-state.

To understand this statement, let us consider some of the properties of the slip velocity field inside a pulse within a pulse train, for both large and small $H$. In the former case, the slip velocity inside the pulse is significantly amplified compared to $v_0$ over a large fraction of the pulse width $w_{\rm p}$, such that most of the pulse is characterized by a slip velocity out of the unstable rate-weakening branch of the friction curve. This is demonstrated in the pulse shown in Fig.~\ref{fig:fig2}b (see Sect.~\ref{subsec:EOM} for additional discussion of the slip velocity amplification). This is qualitatively different from the small $H$ case, $H\!<\!L_{\rm c}\!\ll\!W$; here, the slip velocity is amplified over a scale $H$ that is significantly smaller than $w_{\rm p}$. This situation is closely related to the finite height strip problem of fracture mechanics~\cite{Broberg1999}, where the controlling length scale is the strip height $H$. This implies that a significant fraction of the pulse/interface is characterized by $v(x,t)\!\simeq\!v_0$, and since $w_{\rm p}\!\gg\!L_{\rm c}$ (see Sect.~\ref{subsec:w_p} for additional discussion of the properties of $w_{\rm p}$), this extended region is characterized by a slip velocity belonging to the rate-weakening branch of the friction curve and hence is unstable. Consequently, if $W_{\rm p}\=W$ also for $H\!<\!L_{\rm c}\!\ll\!W$, then we expect to observe repeated instabilities on top of the pulse train. That is, we expect the competition between coarsening and elasto-frictional instabilities to prevent the system from reaching a strict steady-state.

In Fig.~\ref{fig:fig3}a we present a space-time plot (in the same format of Fig.~\ref{fig:fig2}a) obtained using FEM simulations in the $H\!<\!L_{\rm c}\!\ll\!W$ regime. It is important to note that in this regime we do not observe the space-independent stick-slip instability discussed in Sect.~\ref{sec:coarsening} and hence no additional constraints that break translational symmetry along the interface are introduced. The results in Fig.~\ref{fig:fig3}a indicate that coarsening dynamics dominate in this regime as well, leading to a pulse train with $W_{\rm p}\=W$. This pulse train, however, is not in strict steady-state as it is clearly observed to be repeatedly/periodically interrupted by bursts of large slip velocities
(compare to Fig.~\ref{fig:fig2}a, where these bursts are absent and the system reaches a strict steady-state). These observations support the physical picture discussed in the previous paragraph, where coarsening dynamics compete and coexist with repeated elasto-frictional instabilities in the $H\!<\!L_{\rm c}\!\ll\!W$ regime.

In order to better understand these complicated spatiotemporal dynamics, we focus in Fig.~\ref{fig:fig3}b on a sequence of snapshots of $v(x)$ of the pulse within the pulse train observed in the long-time dynamics of Fig.~\ref{fig:fig3}a, during one of the bursts (the snapshots are shown in the pulse train co-moving frame, see figure caption). The snapshots are ordered in time, where the smallest $v(x)$ corresponds to early time and the largest to late time. The time points in which the snapshots were taken are marked by vertical lines in Fig.~\ref{fig:fig3}c (same colors and line styles as in the curves of Fig.~\ref{fig:fig3}b), where the maximal slip velocity $v_{\rm m}$ is plotted as a function of time. The early time field reveals a long plateau, larger than $L_{\rm c}$ (recall that here $L_{\rm c}\!\ll\!W$), featuring $v(x)\!\simeq\!v_0$, as predicted above. Consequently, an elasto-frictional instability is indeed expected.

As is evident from both Fig.~\ref{fig:fig3}b and Fig.~\ref{fig:fig3}c, the slip velocity inside the pulse grows significantly, indicating an elasto-frictional instability. A clear signature of the linear elasto-frictional instability associated with $L_{\rm c}$ is an exponential growth of the slip velocity at the early stages of the instability development. This is indeed demonstrated in the inset of Fig.~\ref{fig:fig3}b, supporting the physical picture discussed above. Moreover, the observed exponential growth rate appears to be in the ballpark of the theoretically estimated growth rate (not shown). Finally, the repeated nature of the instability, already observed in Fig.~\ref{fig:fig3}a, is evident also in Fig.~\ref{fig:fig3}c. There, the maximal slip velocity $v_{\rm m}$ is plotted as a function of time, and it is observed that after the instability shown in Fig.~\ref{fig:fig3}b relaxes, another instability with very similar properties occurs after some interval of time.

The results of this section and of the previous one suggest that velocity-driven frictional systems, at least the rather generic class of frictional systems considered in this paper, are strongly dominated by coarsening dynamics that lead to pulse trains characterized by periodicity determined by system length in the sliding direction. For $H\!<\!L_{\rm c}\!\ll\!W$, the train pulse experiences repeated elasto-frictional instabilities, but for larger $H$'s the pulse train is a stable steady-state solution. It is important to note that in this paper we do not study in depth and quantitatively the coarsening dynamics themselves (this will be done elsewhere), but rather focus on their long-time outcome. Next, we shift our focus to other salient features of the observed pulse trains, i.e.~their propagation velocity $c_{\rm p}$ and the properties of single pulses within the train.

\section{Single pulse properties in the large $H$ limit}
\label{sec:single_pulse}

Self-healing slip pulses, as explained in Sect.~\ref{sec:Intro}, are frictional rupture modes that are believed to be quite prevalent, yet they are not fully understood~\cite{Heaton1990,Zheng1998,Somerville1999,Gabriel2012,Melgar2017,Brener2018}. Our analysis above established that self-healing slip pulses, as part of pulse trains, generically and robustly emerge in velocity-driven frictional sliding dynamics. These results offer a rather unusual opportunity to better understand the physical properties of single pulses, which is the main goal of this section.

A steady-state pulse train is composed of single pulses that repeat themselves with spatial periodicity $W_{\rm p}$, extensively discussed above. Once the train periodicity $W_{\rm p}$ is known --- it was shown above to be coarsening-limited (i.e.~equal to the system length, $W_{\rm p}\=W$) ---, one is interested in the selection of the train velocity $c_{\rm p}$ (which is obviously also the single pulse propagation velocity) and in the spatial distributions of slip velocity $v(x)$ and stress $\tau(x)$ within the single pulse. Obtaining closed-form solutions for the field distributions is a difficult challenge not addressed here; instead, we focus below on the behavior of $v(x)$ and $\tau(x)$ near the leading edge of the single pulse, and on some characteristic properties of $v(x)$ and $\tau(x)$. Mort importantly, as already introduced in Fig.~\ref{fig:fig1}, we are interested in the single pulse width $w_{\rm p}$ and in its average slip velocity $v_{\rm p}$.

The single pulse width $w_{\rm p}$, as illustrated in Fig.~\ref{fig:fig1}c, is defined as the size of the portion of the single pulse for which $v(x)\!\ge\!v_0$. Accordingly, the average slip velocity of a single pulse is defined as $v_{\rm p}\!\equiv\!w_{\rm p}^{-1}\!\int_0^{w_{\rm p}}v(\tilde{x})d\tilde{x}$, where $\tilde{x}\=0$ in this context corresponds to the spatial point on the pulse's leading edge such that $v(\tilde{x}\=0)\=v_0$ (the corresponding relation at the trailing edge reads $v(\tilde{x}\=w_{\rm p})\=v_0$). Our goal in this section is to gain physical insight into the single pulse quantities $c_{\rm p}$, $w_{\rm p}$ and $v_{\rm p}$ as a function of the control parameters $W$ and $v_0$ for the adopted interfacial constitutive law, in the $H\!\to\!\infty$ limit. Moreover, we would like to understand the interrelations between $c_{\rm p}$, $w_{\rm p}$ and $v_{\rm p}$, and clearly three such relations are needed in order to determine these three quantities.

One such relation is readily obtained from the steady-state condition of Eq.~\eqref{eq:avr_v}. The latter implies
\begin{equation}
\label{eq:ss_relation}
\hspace{-0.25cm}W^{-1}\!\left[w_{\rm p}v_{\rm p}+(W-w_{\rm p})v_{\rm s} \right]\!=\!v_0 \quad \!\Longrightarrow\!\quad w_{\rm p}v_{\rm p}\!\simeq\! W v_0 \ ,
\end{equation}
where the slip velocity in the `stick region', $v_{\rm s}$ (cf.~Fig.~\ref{fig:fig1}c), has been assumed to be negligibly small (as is indeed the case). Note that Eq.~\eqref{eq:ss_relation} does not explicitly include the propagation velocity $c_{\rm p}$. In Sect.~\ref{subsec:EOM} we show that another relation between $c_{\rm p}$, $w_{\rm p}$ and $v_{\rm p}$ emerges from the pulse's leading edge behavior, a relation that may be viewed as a pulse equation of motion. A third relation is discussed in Sect.~\ref{subsec:w_p}, where we focus on the pulse width $w_{\rm p}$.

\subsection{Pulse equation of motion: Leading edge singularity, effective fracture energy and propagation velocity}
\label{subsec:EOM}

A slip pulse, like other rupture fronts, is expected to feature nearly singular fields near its leading edge, where the negligibly small slip velocity ahead of the pulse, $v_{\rm s}$, dramatically increases to the peak velocity, $v_{\rm m}$ (cf.~Fig.~\ref{fig:fig1}c), over a short length scale. It has been recently shown~\cite{Barras2019,Barras2020,brener2021unconventional,brener2021theory} that for the class of rate-and-state friction constitutive laws illustrated in Fig.~\ref{fig:fig1}b, where the rate dependence of friction is rather weak, the singular leading edge behavior is described by the classical square root singularity of Linear Elastic Fracture Mechanics (LEFM)~\cite{Freund1998} to a good approximation.

\begin{figure}[ht!]
\includegraphics[width=0.5\textwidth]{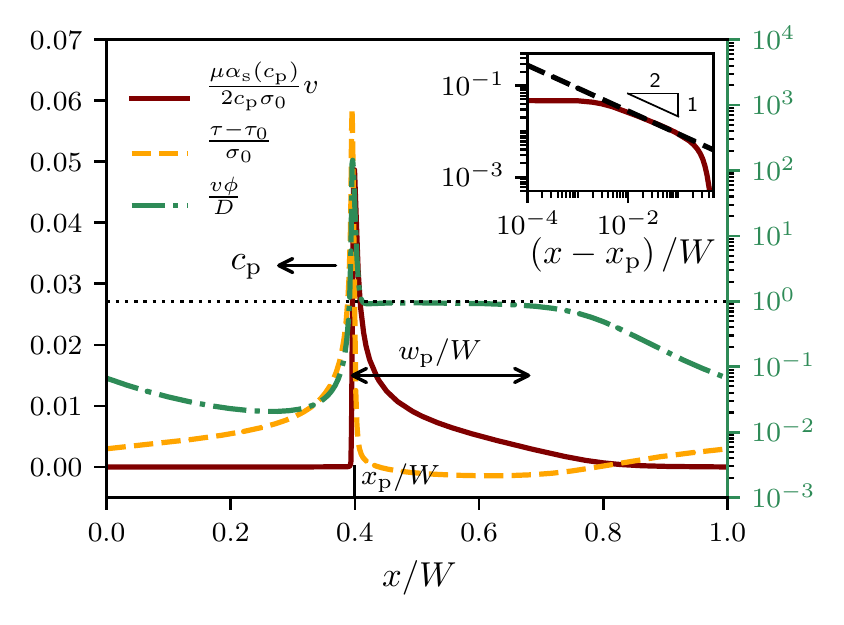}
\caption{The normalized (see legend) profiles of the velocity $v(x)$ (solid brown line, left $y$-axis), the shear/frictional stress $\tau(x)$ (dashed orange, left $y$-axis) and $v(x)\phi(x)/D$ (dashed-dotted green line, right $y$-axis) of a steady-state pulse train propagating to the left at $c_{\rm p}\!=\!0.85 c_{\rm s}$, as obtained by a BIM simulation ($H\!\to\!\infty$) with $v_0\!=\!3\!\times\!10^{-3} \text{m/s}$, $W\!=\!14.07 \text{m}$ and the N-shaped steady-state friction law of Fig.~\ref{fig:fig1}b. The leading edge is indicated by $x_{\rm p}/W$ and the pulse width by $w_{\rm p}/W$ (and a double-arrow). Both the stress and the velocity fields reveal a singular behavior near the pulse leading edge, see text for discussion. (inset) The velocity field $v(x)$ (solid brown line) behind the leading edge, on a double-logarithmic scale, with the $x$ axis being $(x-x_{\rm p})/W$. The dashed black line features a slope of $-1/2$, characteristic of the square root singularity of LEFM, see text for additional discussion.}
\label{fig:fig4}
\end{figure}

In the framework of LEFM, the slip velocity behind the leading pulse edge is expected to take the following universal form~\cite{Freund1998}
\begin{equation}
\label{eq:sqrt}
v(x)\simeq\frac{2\,c_{\rm p}\,K_{\rm III}}{\alpha_{\rm s}(c_{\rm p})\,\mu\,\sqrt{2\pi (x-x_{\rm p})}} \ ,
\end{equation}
where $\alpha_{\rm s}(c_{\rm p})\=\sqrt{1-(c_{\rm p}/c_s)^2}$, $x_{\rm p}$ is the leading edge position and $K_{\rm III}$ is the mode-III stress intensity factor, quantifying the amplitude of the universal square root singularity (Eq.~\eqref{eq:sqrt} refers to mode-III conditions, which we focus on in this section, but the same singularity applies to mode-II conditions~\cite{Freund1998}). In Fig.~\ref{fig:fig4} we plot $v(x)$ (properly normalized, see legend) over the entire pulse train period $W_{\rm p}\=W$ (solid brown line, left $y$-axis). To test whether the observed amplified slip velocity near the leading edge indeed follows the square root singularity of Eq.~\eqref{eq:sqrt}, we plot in the inset the same normalized $v(x)$ behind the leading edge against $(x-x_{\rm p})/W$ on a double-logarithmic scale. As is indicated by the dashed line of slope $-\tfrac{1}{2}$, there exists a spatial range near the leading edge over which $v(x)$ follows the square root singular behavior of Eq.~\eqref{eq:sqrt}.

The singular slip velocity behavior of Eq.~\eqref{eq:sqrt} behind the pulse's leading edge is accompanied by a shear/frictional stress field $\tau(x)$ that features the same square root singularity \emph{ahead} of the leading edge and a predominantly constant residual stress behind it~\cite{Barras2019,Barras2020}. In Fig.~\ref{fig:fig4} we present $\tau(x)$ (properly normalized, see legend) again over the entire pulse train period $W_{\rm p}\=W$ (dashed orange line, sharing the \emph{same} left $y$-axis with the normalized $v(x)$ field). It is observed that indeed $\tau(x)$ features significant amplification ahead of the leading edge (its quantification is not presented here) and that it is approximately constant behind the leading edge inside the pulse. Finally, we present in Fig.~\ref{fig:fig4} $v(x)\phi(x)/D$ (dashed-dotted green line, right $y$-axis), whose deviation from unity provides a measure of the degree by which $\phi(x)$ is out of steady-state with respect to $v(x)$. It is observed that inside the pulse $v(x)\phi(x)/D\!\simeq\!1$ (see horizontal dashed line) almost everywhere. That is, inside the pulse away from the leading edge transition region, $\phi(x)$ is ``equilibrated'' with $v(x)$.

The approximate validity of Eq.~\eqref{eq:sqrt} has significant implications for the pulse dynamics. Most notably, the square root singular fields are accompanied by a finite (non-singular) flux of energy $G$ into the leading edge region. The latter is dissipated near the leading edge, i.e.~it is balanced by an effective fracture energy $G_{\rm c}$, which quantifies the dissipation involved in slip pulse propagation (on top of the background frictional dissipation~\cite{kanamori2000microscopic,Barras2020}). That is, we expect the following relation~\cite{Freund1998}
\begin{equation}
\label{eq:EOM}
G_{\rm c} \simeq G = \frac{K^2_{\rm III}}{2\alpha_{\rm s}(c_{\rm p})\,\mu} \ ,
\end{equation}
to approximately hold. The effective fracture energy $G_{\rm c}$ can be independently estimated using the interfacial constitutive law, following~\cite{Rubin2005,Ampuero2008} and as shown in Appendix~\ref{app:Gc}. With $G_{\rm c}$ at hand, one can eliminate $K_{\rm III}$ between Eqs.~\eqref{eq:sqrt}-\eqref{eq:EOM}, yielding an expression for the singular part of $v(x)$ near the leading edge of the pulse.

The resulting $v(x)$ allows to derive a relation between $c_{\rm p}$, $w_{\rm p}$ and $v_{\rm p}$. It is important to note that the latter two quantities characterize the whole pulse, not just its singular part. To bridge over this gap, we assume that the singular part of $v(x)$ describes reasonably well the entire pulse, up to an overall shift that dominates the trailing edge behavior. The shift can be determined by demanding $v(x_{\rm p}\!+\!w_{\rm p})\=v_0$, which follows from the definition of $w_{\rm p}$, leading to
\begin{equation}
\label{eq:v_over_pulse}
\frac{v(x)}{c_s}\!\simeq\!\sqrt{\frac{4G_c}{\pi\mu}}\frac{c_p}{c_s\sqrt{\alpha_s(c_p)}} \left(\frac{1}{\sqrt{x-x_{\rm p}}}\!-\!\frac{1}{\sqrt{w_p}}\right)+\frac{v_0}{c_s} \ .
\end{equation}
With this approximation at hand, we can then calculate $v_{\rm p}$ according to its definition $v_{\rm p}\=w_{\rm p}^{-1}\!\int_{x_{\rm p}}^{x_{\rm p}+w_{\rm p}}v(\tilde{x})d\tilde{x}$, leading to
\begin{equation}
\label{eq:eom_prediction}
\frac{c_{\rm p}}{c_s\sqrt{\alpha_s(c_{\rm p})}} \sim \sqrt{\frac{\pi\,\mu\, w_{\rm p}}{G_{\rm c}}} \,\frac{v_{\rm p}-v_0}{2c_s} \ .
\end{equation}

\begin{figure}[ht!]
\includegraphics[width=0.5\textwidth]{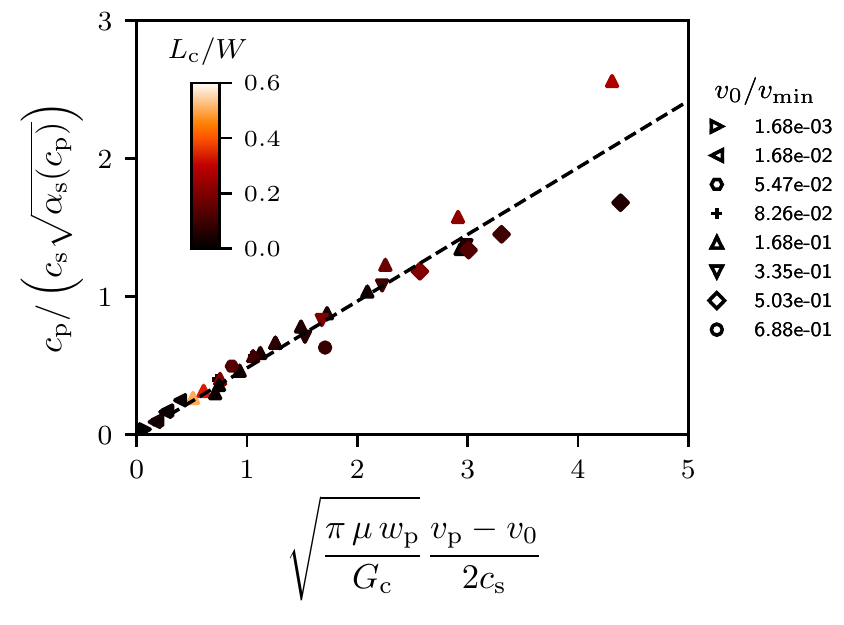}
\caption{Testing the theoretical prediction of Eq.~\eqref{eq:eom_prediction} for a broad range of parameters, including variations of the system length $W$, the driving velocity $v_0$, the shear modulus $\mu$ and the characteristic slip displacement $D$, see Appendix~\ref{app:friction_laws} and~\ref{app:BIM} for additional details and the exact values used. The data correspond to BIM simulations with the N-shaped friction law of Fig.~\ref{fig:fig1}b, revealing very good agreement with the prediction, with an ${\cal O}(1)$ pre-factor. $G_{\rm c}$ has been estimated following~\cite{Rubin2005,Ampuero2008}, see Appendix~\ref{app:Gc} for more details.}
\label{fig:fig5}
\end{figure}

The prediction in Eq.~\eqref{eq:eom_prediction} is a relation between $c_{\rm p}$ (left hand side), and $w_{\rm p}$ and $v_{\rm p}$ (right hand side), where the pre-factor is expected to be ${\cal O}(1)$. It may be viewed as an equation of motion for the pulse. Moreover, note that Eq.~\eqref{eq:ss_relation} (which has been independently verified) can be used to eliminate $v_{\rm p}$ from the right hand side of Eq.~\eqref{eq:eom_prediction}, making it a function of $w_{\rm p}$ alone. Equation~\eqref{eq:eom_prediction} is tested in Fig.~\ref{fig:fig5} over a broad range of $W$ and $v_0$ values. It is observed that the theoretical prediction is strongly supported by the simulational data, where the pre-factor is indeed ${\cal O}(1)$. Consequently, only one additional relation between $c_{\rm p}$, $w_{\rm p}$ and $v_{\rm p}$ is needed in order to fully determine the single pulse properties. In particular, predicting the pulse width $w_{\rm p}$ in terms of $W$ and $v_0$ would be sufficient since Eqs.~\eqref{eq:ss_relation} and~\eqref{eq:eom_prediction} would then allow to calculate $v_{\rm p}$ and $c_{\rm p}$, respectively. Therefore, we next consider $w_{\rm p}(W,v_0)$.

\subsection{The pulse width}
\label{subsec:w_p}

Our goal in this subsection is to discuss the pulse width $w_{\rm p}(W,v_0)$. It would be useful in this context, as a preparatory step, to consider the ``ideal'' solution for an isolated pulse. The latter corresponds to an isolated pulse in an infinite medium ($H\!\to\!\infty$) whose slip velocity $v^{(i)}(x)$ (the superscript $(i)$ denotes hereafter ``ideal pulse'') vanishes for $x\!<\!x_{\rm p}$ and $x\!>\!2w^{(i)}_{\rm p}+x_{\rm p}$, and the frictional strength is constant for $x_{\rm p}\!\le\!x\!\le\!2w^{(i)}_{\rm p}+x_{\rm p}$. The solution takes the form~\cite{Kostrov1964,freund1979mechanics,Broberg1999}
\begin{equation}
\label{eq:ideal_pulse}
v^{(i)}(x)=v_0 \sqrt{\frac{2w^{(i)}_{\rm p}+x_{\rm p}-x}{x-x_{\rm p}}}
\end{equation}
for $x_{\rm p}\!\le\!x\!\le\!2w^{(i)}_{\rm p}+x_{\rm p}$. Note that the solution in Eq.~\eqref{eq:ideal_pulse} features the square root singularity of Eq.~\eqref{eq:sqrt} in the $x\!\to\!x_{\rm p}^+$ limit and it is consistent with the single pulse width definition adopted throughout this paper. That is, we have $v^{(i)}(x\=w^{(i)}_{\rm p}\!+\!x_{\rm p})\=v_0$ such that $w^{(i)}_{\rm p}$ retains its meaning as the pulse width.

The physics underlying the ideal pulse solution in Eq.~\eqref{eq:ideal_pulse} differs from the physics underlying the train pulse solutions derived above in three major respects: (i) Our solutions correspond to rate-and-state friction, while Eq.~\eqref{eq:ideal_pulse} corresponds to a \emph{constant} frictional strength inside the pulse. (ii) Eq.~\eqref{eq:ideal_pulse} corresponds to an isolated pulse, i.e.~to a single pulse that does not interact with other pulses in the train over the periodicity scale $W_{\rm p}\=W$. (iii) $w^{(i)}_{\rm p}$ in Eq.~\eqref{eq:ideal_pulse} is not selected, an issue that is obviously related to points (i) and (ii).

An attempt to remedy points (ii) and (iii) above, which is inevitably \emph{superficial}, would be to impose the steady-state pulse train condition of Eq.~\eqref{eq:avr_v} on Eq.~\eqref{eq:ideal_pulse}, resulting in $w^{(i)}_{\rm p}/W\=1/\pi$. That is, we construct an approximate pulse train solution by concatenating/superimposing the ideal pulse solutions of Eq.~\eqref{eq:ideal_pulse} with periodicity $W_{\rm p}\=W$ and a single pulse width $w^{(i)}_{\rm p}/W\=1/\pi$. Below we show that this artificially constructed pulse train qualitatively fails to describe our rate-and-state friction pulse train solutions, where $w_{\rm p}/W$ is not a constant, but rather depends on both $W$ and $v_0$. Moreover, while both the ideal pulse solution of Eq.~\eqref{eq:ideal_pulse} and our pulses (cf.~Figs.~\ref{fig:fig2}b and~\ref{fig:fig4}) do not feature a singularity at the trailing edge, the pulse shape $v(x)$ near the trailing edge differs quite significantly (not shown). Another difference between the two solutions is that while the ideal pulse solution is truly singular at the leading edge, the singularity in our solutions is self-consistently regularized on small scales near the leading edge.

In light of the discussion of the ideal pulse above, which mainly served to highlight the differences compared to our pulse train solutions, our next goal is to better characterize the function $w_{\rm p}(W,v_0)/W$. The first question we need to address is how to properly nondimensionalize $W$ and $v_0$, i.e.~the dimensionless pulse width $w_{\rm p}/W$ should be expressed in terms of dimensionless quantities. We suggest that the proper way to nondimensionalize $W$ and $v_0$ is through their relation to the elasto-frictional instability, which is a necessary condition for the emergence of pulse trains.
\begin{figure}[ht!]
\includegraphics[width=0.5\textwidth]{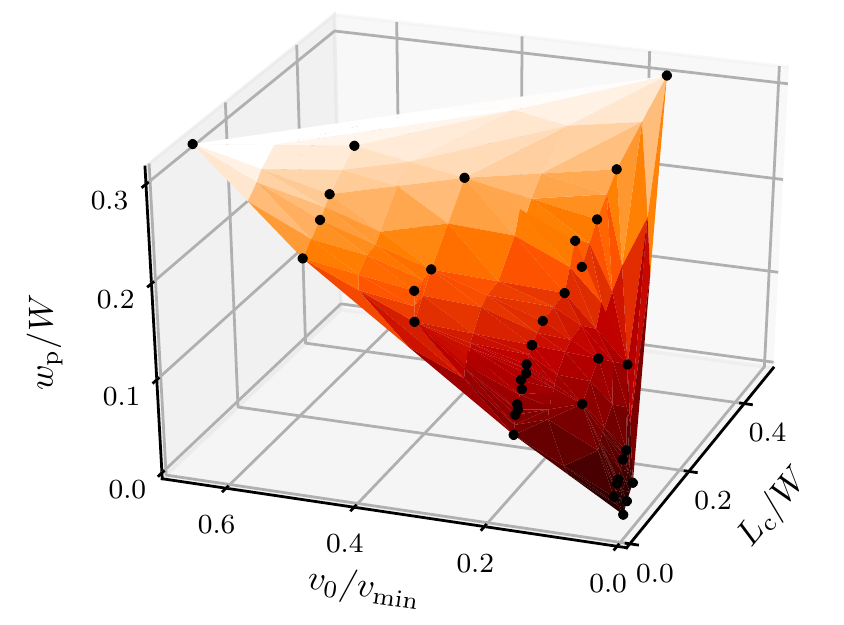}
\caption{The dimensionless pulse width $w_{\rm p}/W$ as a function of both the normalized driving velocity $v_0/v_{\rm min}$ and system size $L_{\rm c}/W$, see text for additional discussion. Each black circle corresponds to one of the simulations shown in Fig.~\ref{fig:fig5}, spanning a broad range of system parameters (including variations of the system size $W$, driving velocity $v_0$, shear modulus $\mu$ and the characteristic slip displacement $D$). Variations of $\mu$ alter the elasto-frictional length $L_{\rm c}$, while variations of $D$ affect both $L_{\rm c}$ and $v_{\rm min}$. The surface is interpolated from the data points, supporting the existence of  a smooth and monotonically increasing function $w_{\rm p}/W\!=\!{\cal G}(L_{\rm c}/W,v_0/v_{\rm min})$, as predicted theoretically (see text for discussion).}
\label{fig:fig6}
\end{figure}

The elasto-frictional instability exists if the system is driven at a velocity $v_0$ that resides on the rate-weakening branch of the steady-state friction curve $f_{\rm ss}(v)$ (cf.~Fig.~\ref{fig:fig1}b), i.e.~when $v_0$ is larger than the slip velocity at which $f_{\rm ss}(v)$ attains its maximum and smaller than $v_{\rm min}$. In the limit $v_0\!\to\!v_{\rm min}$, the system will approach a stable homogeneous sliding state; that is, as $v_0/v_{\rm min}$ is increased toward unity, we expect the pulse width to approach $W$ (and obviously its slip velocity to approach $v_0$). For $v_0\!<\!v_{\rm min}$, the elasto-frictional instability manifests itself only for $L_{\rm c}\!<\!W$, i.e.~when an unstable perturbation can fit into the system. Consequently, in the limit $W\!\to\!L_{\rm c}^+$ the system will approach a stable homogeneous sliding state; that is, as $L_{\rm c}/W$ is increased toward unity, we expect the pulse width to approach $W$.

Taken together, these physical considerations lead us to expect that there exists a smooth function $w_{\rm p}/W\={\cal G}(L_{\rm c}/W,v_0/v_{\rm min})$ that monotonically increases with its two arguments. This prediction is tested in Fig.~\ref{fig:fig6}, where $w_{\rm p}/W$ is plotted against $L_{\rm c}/W$ and $v_0/v_{\rm min}$. It is observed that a smooth $w_{\rm p}/W\={\cal G}(L_{\rm c}/W,v_0/v_{\rm min})$ function seems to exist, and that $w_{\rm p}/W$ indeed increases with both of its arguments. The theoretical derivation of the precise functional form of $w_{\rm p}/W\={\cal G}(L_{\rm c}/W,v_0/v_{\rm min})$ is left as a challenge for future research. With this analysis, we conclude our discussion of the single pulse properties $c_{\rm p}$, $w_{\rm p}$ and $v_{\rm p}$ in the large $H$ limit.

\section{Summary and Discussion}
\label{sec:summary}

In this work we extensively studied velocity-driven frictional systems, using a combination of computational and theoretical approaches. Velocity-driven frictional systems exist in a broad range of engineering and tribological applications, as well as in geophysical contexts. We show that such frictional systems, described within the experimentally motivated rate-and-state friction constitutive framework, give rise to the emergence of pulse trains once driven at a sliding velocity for which the frictional interface is rate-weakening. The propagating pulse train is a non-equilibrium dissipative analog of equilibrium phase separation in thermodynamic systems. We find that such velocity-driven frictional systems undergo coarsening dynamics leading to train periodicity set by the size of the system in the sliding direction, independently of the height of the system. The properties of single pulses within the train are quantitatively and comprehensively analyzed and interpreted. In particular, the pulse propagation velocity, width and average slip rate are shown to be related through an equation of motion that is associated with the nearly singular fields in the vicinity of the leading pulse edge.

Throughout this work we employed periodic boundary conditions in the sliding direction. Similar periodic boundary conditions have been employed in the velocity-driven frictional simulations of~\cite{Heimisson2019}, where instability has been induced by various physical processes leading to a reduction in the effective normal stress, rather than by rate-weakening friction per se (note that the possible emergence of pulse trains under rate-strengthening friction has been discussed in~\cite{Brener2005}). While coarsening dynamics have not been discussed in~\cite{Heimisson2019}, the results therein (cf.~Fig.~4 in~\cite{Heimisson2019}) do indicate similar coarsening dynamics as found here. Such periodic boundary conditions naturally emerge in annular/rotary shear geometry experiments~\cite{galeano2000experimental,ma2014rotary}, which is also characteristic of many engineering systems. Such an experimental setup offers a natural test bed for our predictions. Moreover, we believe that the coarsening dynamics revealed in this work are physically relevant for sufficiently long systems (large $W$) that do not feature periodic boundary conditions, as well as for geophysical fault dynamics, issues that should be addressed in future work.

This work offers a rather robust framework for studying self-healing pulses, which are of importance in many frictional systems. Our results also pose several interesting questions. First, calculating the single pulse shape and consequently its width $w_{\rm p}$ (cf.~Fig.~\ref{fig:fig6}) remain open challenges. Second, we have not studied in any detail the coarsening dynamics themselves, but rather focused on their long-time outcome. Yet, earlier time coarsening (cf.~Fig.~\ref{fig:fig2}a) may be characterized by rather complex spatiotemporal dynamics, possibly chaotic ones. Quantitatively analyzing these dynamics and their acoustic emission signature may be important for understanding the non-steady frictional response of various physical systems. Moreover, in an even broader statistical physics context, such complex dynamics may reveal how anomalous statistical properties (e.g.~fat power-law statistical distributions~\cite{rundle2003statistical,kawamura2012statistical}) spontaneously emerge even in the absence of input disorder (quenched or thermal). These interesting issues will be addressed in future investigations.\\

{\em Acknowledgements} E.~B.~acknowledges support from the Israel Science Foundation (Grant no.~1085/20), the Ben May Center for Chemical Theory and Computation and the Harold Perlman Family. J.-F.~M.~and T.~R.~acknowledge internal support from EPFL.

\setcounter{equation}{0}
\setcounter{figure}{0}
\setcounter{section}{0}
\setcounter{table}{0}
\makeatletter
\renewcommand{\thefigure}{S\arabic{figure}}

\vspace{0.6cm}

{\large \bf Appendices}

\appendix

\section{The interfacial constitutive law}
\label{app:friction_laws}

\begin{figure}[ht!]
\includegraphics[width=0.5\textwidth]{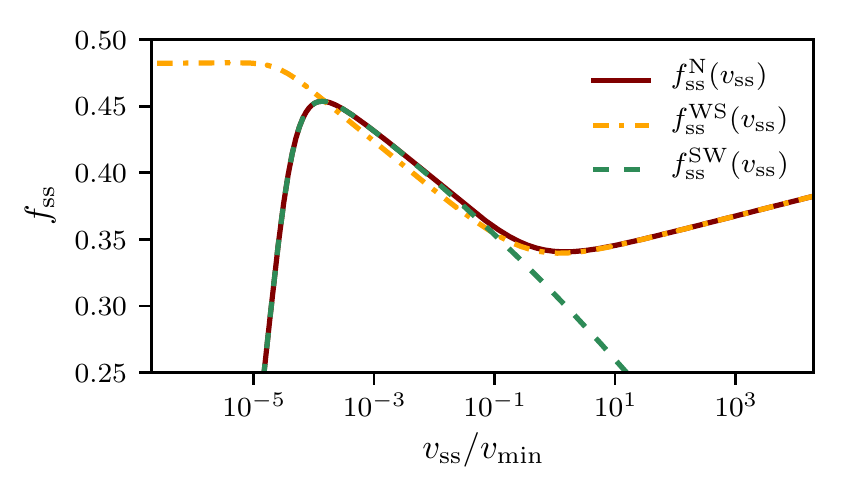}
\caption{The normalized steady-state frictional strength $f_{\rm ss}$ vs.~the steady-state normalized slip velocity $v_{\rm ss}$, presented on a semi-logarithmic scale for three constitutive laws. The first (solid brown curve), which features an N shape and is denoted by $f^{\rm N}_{\rm ss}(v_{\rm ss})$ (see legend), corresponds to Eqs.~\eqref{eq:n_shape}-\eqref{eq:evolution_law} and has already been presented in Fig.~\ref{fig:fig1}b. The local minimum of $f^{\rm N}_{\rm ss}(v_{\rm ss})$, denoted by $v_{\rm min}$, is used to normalize $v_{\rm ss}$. The second constitutive law (dotted-dashed orange curve) corresponds to Eqs.~\eqref{eq:evolution_law}-\eqref{eq:ws_law} and is denoted by $f^{\rm WS}_{\rm ss}(v_{\rm ss})$, see text for additional discussion. Finally, the third constitutive law (dashed green curve) features no local minimum (see text for discussion) and is denoted by $f^{\rm SW}_{\rm ss}(v_{\rm ss})$. All three constitutive laws share the same rate-weakening behavior of the steady-state frictional strength at intermediate slip velocities, which plays a major role in this work.}
\label{fig:figSM1}
\end{figure}

The rate-and-state friction constitutive framework~\cite{Baumberger2006,Ruina1983,Marone1998a,Nakatani2001} has been formulated in Eqs.~\eqref{eq:friction_law}-\eqref{eq:dot_phi}. Here we specify the constitutive functions $f(\cdot)$ and $g(\cdot)$, which appear in these two equations respectively, taking the explicit forms~\cite{Baumberger2006,Ruina1983,Marone1998a,Nakatani2001,Bhattacharya2014,Bar-Sinai2012,Bar-Sinai2014,Bar-Sinai2015,Bar-Sinai2019,Aldam2017,Brener2018,Barras2019}
\begin{equation}
\begin{split}
\label{eq:n_shape}
f^{\rm N}(|v|,\phi)  &=  \left[ 1 + b \log \left( 1 + \frac{\phi}{\phi_*} \right) \right] \\
 & \times \left[ \frac{f_0}{\sqrt{1+(v_*/|v|)^2}} + a \log \left(1+\frac{|v|}{v_*}\right) \right]
\end{split}
\end{equation}
and
\begin{equation}
\label{eq:evolution_law}
g(|v|,\phi) = 1  - \frac{|v|\phi}{D}\sqrt{1 + (v_* / v)^2} \ .
\end{equation}
These functions feature 6 parameters: $f_0$, $a$, $b$, $D$, $v_*$ and $\phi_*$, whose values are specified in Table~\ref{tab:frictionParameters}. The superscript `N' in Eq.~\eqref{eq:n_shape} denotes the fact that the steady-state friction curve $f_{\rm ss}(v)$ corresponding to Eqs.~\eqref{eq:n_shape}-\eqref{eq:evolution_law} features an N shape, as shown in Fig.~\ref{fig:fig1}b and again here in
Fig.~\ref{fig:figSM1} (solid brown curve).

The N-shaped steady-state friction curve features rate-strengthening at very low velocities, rate-weakening at intermediate velocities and again rate-strengthening at high velocities, i.e.~it follows a strengthening-weakening-strengthening (SWS) sequence with increasing slip velocities. As highlighted in the text, the most crucial branch for the emergence of pulse trains is the rate-weakening one. Consequently, we also considered two variants of SWS-related function $f^{\rm N}(|v|,\phi)$ of Eq.~\eqref{eq:n_shape}. One variant corresponds to replacing Eq.~\eqref{eq:n_shape} with
\begin{equation}
\label{eq:ws_law}
f^{\rm WS}(|v|,\phi)\!=\! f_0 \!\left[1\!+\!b \log \left(\!1\!+\!\frac{\phi}{\phi_*}\!\right)\!\right] \!+\! a \log\left(\!1\!+\!\frac{|v|}{v_*}\!\right) \ .
\end{equation}
The resulting steady-state curve is rate-independent at very low velocities, rate-weakening at intermediate velocities and rate-strengthening at high velocities. Consequently, we use the superscript `WS' for this function and plot the corresponding steady-state curve in Fig.~\ref{fig:figSM1} (dotted-dashed orange curve).

The second variant corresponds to omitting the `+1' in the logarithm that multiplies $b$ in the first square brackets on the right hand side of Eq.~\eqref{eq:n_shape}. This modification eliminates the high velocities rate-strengthening branch of the steady-state friction curve~\cite{Bar-Sinai2014}. Consequently, we term the resulting $f(\cdot)$ that replaces Eq.~\eqref{eq:n_shape} $f^{\rm SW}(|v|,\phi)$ and plot the corresponding steady-state curve in Fig.~\ref{fig:figSM1} (dashed green curve). All three steady-state curves, corresponding to $f^{\rm N}(|v|,\phi)$, $f^{\rm WS}(|v|,\phi)$ and $f^{\rm SW}(|v|,\phi)$, essentially share the same rate-weakening behavior at intermediate slip velocities. Results for $f^{\rm WS}(|v|,\phi)$ and $f^{\rm SW}(|v|,\phi)$ are presented below.

\begin{table}[ht!]
  \begin{center}
    \begin{tabular}{p{3cm} p{3cm} r}
      \hline
      \hline
      Parameter & Value & Unit \\
      \hline
      $f_0$ & 0.28 & ... \\
      $a$ & 0.005 & ... \\
      $b$ & 0.075 & ... \\
      $D$ & 5 $\times$ 10\textsuperscript{-7} & m \\
      $v_{*}$ & 1 $\times$ 10\textsuperscript{-7} & m/s \\
      $\phi_{*}$ & 3.3 $\times$ 10\textsuperscript{-4} & s \\
      \hline
      \hline
    \end{tabular}
  \end{center}
  \caption{Typical values of rate-and-state parameters used in this work. In addition, $D$ has been varied from $1.25\!\times\!10$\textsuperscript{-7} to $1.5\!\times\!10$\textsuperscript{-6}m in the results appearing in Figs.~\ref{fig:fig5},~\ref{fig:fig6},~\ref{fig:figSM3}.
    \label{tab:frictionParameters}}
\end{table}

\section{The elasto-frictional length $L_{\rm c}$}
\label{app:nucleation_length}

The calculation of the elasto-frictional length $L_{\rm c}$ associated with the rate-weakening instability appears in previous works~\cite{Rice1983,Ruina1983,Lapusta2000,Baumberger2006,Bhattacharya2014,Aldam2017,Bar-Sinai2019}, so here we just very briefly highlight the structure of the calculation. A linear stability analysis of an interface sliding at a constant velocity $v_0$ is performed; that is, the starting point is a space and time independent solution featuring a steady-state frictional stress $\tau\=\sigma_0 f_{\rm ss}(v_0)$. One then introduces small perturbations to all fields, each assumed to be proportional to a Fourier mode $e^{\Lambda t - i k x}$, where $\Lambda$ is the complex growth rate and $k$ is the wavenumber. Obtaining expressions for the perturbation of the interfacial shear stress, from the bulk elastodynamic equations corresponding to bodies of height $H$, and for the perturbation of the frictional strength, one obtains (by equating the two) the linear perturbation spectrum $\Lambda(k)$. The elasto-frictional length is related to the critical wavenumber $k_{\rm c}$ for the onset of instability according to $L_{\rm c}(H)\=2\pi / k_{\rm c}$, where $k_{\rm c}$ is determined from the zero crossing of the real part of $\Lambda$, i.e.~by $\Re[\Lambda(k_{\rm c},H)]\=0$. The general structure of $L_{\rm c}(H)$ appears in Eq.~\eqref{eq:Lc}.

\section{The spectral boundary integral formulation}
\label{app:BIM}

The infinite height ($H\!\to\!\infty$) calculations are performed using an in-house open-source implementation (called cRacklet~\cite{cRacklet}) of the spectral boundary integral formulation of the elastodynamic equations~\cite{Geubelle1995,Morrissey1997,Breitenfeld1998}. The basic relation between the interfacial shear stress and the slip displacement in this case is given in Eq.~\eqref{eq:BIM}, where the Fourier representation of the spatiotemporal integral term $s(x,t)$ for both mode-II and mode-III symmetries can be found in~\cite{Breitenfeld1998}. In our numerical calculations, the interface is assumed to be initially at steady-state with $v_{\rm ss}\=v_0$ and $\phi_{ss}\=D/v_0$. The interface is then perturbed by adding spatial Gaussian noise to the state variable $\phi(x,t)$, and the resulting slip velocity is computed by combining Eq.~\eqref{eq:BIM} and the rate-and-state friction law $\tau\=\sigma_0\sgn(v) f(|v|,\phi)$. Note that $\tau_0(t)$ in Eq.~\eqref{eq:BIM} is treated as unknown and that we impose the constraint of Eq.~\eqref{eq:avr_v}. The slip displacement $u(x,t)$ is then integrated in time using an explicit time-stepping scheme $u(x,t+\Delta t)\= u(x,t) + \tfrac{1}{2} v(x,t)\Delta{t}$, with the time step being $\Delta{t}\= \alpha_{_{\rm BIM}}\Delta{x}/ c_s$, where $\Delta{x}$ is the numerical grid spacing. The numerical parameter $\alpha_{_{\rm BIM}}$ is chosen to ensure the stability and the convergence of the numerical scheme, and is typically set equal to $0.1$.

The exact shape of the initial perturbation has no impact on the long-time behavior of the system (i.e.~it does not alter the properties of the emergent steady pulse trains). The coarsening dynamics, which are not studied in detail in this work, may depend on the initial perturbation. The bulk parameters in our calculations have been set equal to $\mu\=9\!\times\!10$\textsuperscript{9}Pa, $\nu\=0.33$ and $\rho\=1200$kg/m\textsuperscript{3}. In addition, $\mu$ has been varied from $2.25\!\times\!10$\textsuperscript{9} to $9\!\times\!10$\textsuperscript{10}Pa in the results appearing in Figs.~\ref{fig:fig5},~\ref{fig:fig6},~\ref{fig:figSM3}.

\section{Finite element method formulation}
\label{app:FEM}

\subsection{The numerical method}

The finite height $H$ calculations follow the system configuration sketched in Fig.~\ref{fig:fig1}a. The top and bottom boundaries are loaded by a horizontal velocity $v_0/2$ (in opposite directions, cf.~Fig.~\ref{fig:fig1}a) and by a constant compressive normal stress of magnitude $\sigma_0$. The bodies are initially moving uniformly in opposite directions at a velocity $v_0/2$, and periodic boundary conditions are enforced at the lateral edges, $x\=0$ and $x\=W$. The interface is initially at steady-state with $v_{\rm ss}\=v_0$ and $\phi_{ss}\=D/v_0$. Perturbations are introduced by adding spatial Gaussian noise to the state variable, as in the BIM case. The bulk parameters used for the FEM simulations and their BIM counterparts (see, for example, Fig.~\ref{fig:fig2}b) are set equal to $\mu\=3.1\!\times\!10$\textsuperscript{9}Pa, $\nu\=0.33$ and $\rho\=1200$kg/m\textsuperscript{3}.

The FEM calculations are performed using the explicit dynamic finite element framework, based on an in-house open-source finite element library called Akantu~\cite{richart2015}. The domain is discretized into a regular mesh composed of bilinear quadrilateral elements (Q4). The sliding interface between the two elastic bodies is modeled using a node to node contact algorithm, whose details can be found in~\cite{Rezakhani2020}. Time is integrated using the central difference method and the time step is taken small enough to eliminate the numerical instabilities associated with the explicit finite element modeling of rate-and-state friction, as explained in~\cite{Rezakhani2020}. In our simulations, we set the time step to $\Delta{t}\=\alpha_{_{\rm FEM}} \Delta{t}_{_{\rm CFL}}$, where $\Delta{t}_{_{\rm CFL}}$ is set by the Courant-Friedrichs-Lewy condition and $\alpha_{_{\rm FEM}}$ is typically taken to be ${\cal O}(0.01)$.

\subsection{The WS friction law}
\label{app:ws_friction}

\begin{figure}[ht!]
\includegraphics[width=0.5\textwidth]{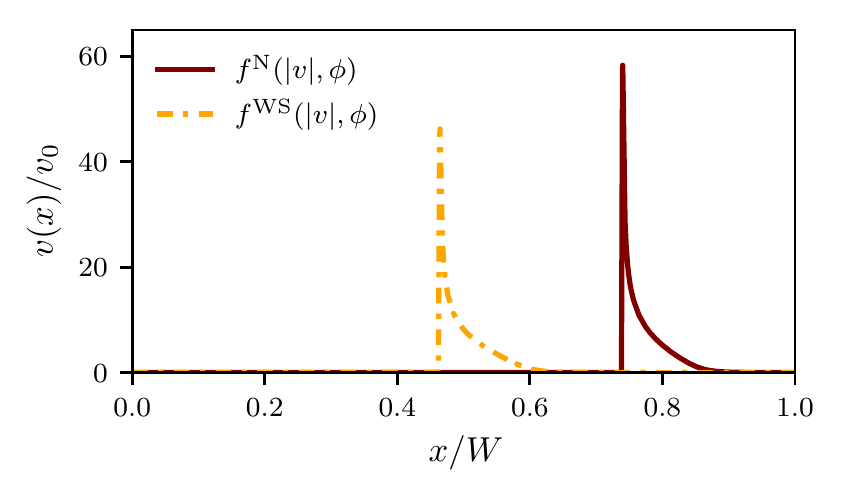}
\caption{Snapshots of the steady-state slip velocity field $v(x)/v_0$ obtained in two BIM simulations, one with the N-shaped friction law (Eq.~\eqref{eq:n_shape}, solid brown curve, cf.~Fig.~\ref{fig:figSM1}) and the other with the WS friction law (Eq.~\eqref{eq:ws_law}, dotted-dashed orange curve, cf.~Fig.~\ref{fig:figSM1}). The two simulations are otherwise identical, employing $v_0\!=\!1\!\times\!10^{-3}\text{m/s}$ and $W\!=\!40\text{m}$. In both cases, the resulting pulse train periodicity satisfies $W_{\rm p}\!=\!W$, the single pulse shape is quantitatively similar (note that the two snapshots are shifted along $x$ for visual clarity) and the propagation velocity is also quantitatively similar ($c_{\rm p}^{\rm N} \!\simeq\!0.73c_s$ for the N-shaped law and $c_{\rm p}^{\rm WS}\!\simeq\!0.70c_s$ for the WS law).}
\label{fig:figSM2}
\end{figure}

As explained in the main text and in Appendix~\ref{app:friction_laws}, the most important aspect of the interfacial constitutive relation for our results is the existence of a rate-weakening branch of the steady-state friction curve and that the applied velocity $v_0$ resides on it. To demonstrate this point, we performed exactly the same BIM simulations except that in one case we used $f^{\rm N}(|v|,\phi)$ of Eq.~\eqref{eq:n_shape} and in the other $f^{\rm WS}(|v|,\phi)$ of Eq.~\eqref{eq:ws_law}, which are mainly distinguished by the different steady-state behavior at very low velocities, see Fig.~\ref{fig:figSM1}. The long-time behavior of the system in the two cases is presented in Fig.~\ref{fig:figSM2} (employing the same line type and color scheme of Fig.~\ref{fig:figSM1}). It is observed that the results are quantitatively similar, in terms of the train periodicity ($W_{\rm p}\=W$), single pulse shape and propagation velocity (see figure legend), substantiating our claim.

It turned out that this robustness of our results against changes in the low velocity behavior of the friction law (while preserving a very similar rate-weakening behavior) is of technical importance in relation to our FEM calculations. In particular, we found that the FEM calculations suffered from numerical instabilities associated with the low frictional strength featured by the N-shaped interfacial constitutive relation at low velocities. These numerical instabilities can be eliminated by using Eq.~\eqref{eq:ws_law} instead of Eq.~\eqref{eq:n_shape}, without affecting the emerging physics. Consequently, in our FEM calculations we adopted the WS friction law of Eq.~\eqref{eq:ws_law}.

\subsection{The space-independent stick-slip like behavior for $H\!\simeq\!W\!\gg\!L_{\rm c}$}
\label{app:mode_0}

As explained in the main text, our FEM calculations with $H\!\simeq\!W\!\gg\!L_{\rm c}$ featured time periods where the slip velocity almost vanished homogeneously across the interface, followed by nearly homogeneous large slip velocity periods. This space-independent stick-slip like behavior appears to substantially deviate from the steady-average velocity condition of Eq.~\eqref{eq:avr_v}, and hence is not expected to persist in the large $W$ and long-time limits. Consequently, we slightly modified our numerical scheme to eliminate this behavior by adding a constraint in the FEM $H\!\simeq\!W\!\gg\!L_{\rm c}$ simulations. In particular, at each time step we computed the average slip rate $\langle v\rangle$ of the interface and compared it to the driving velocity $v_0$. We then introduced a small spatially-homogeneous shift $\Delta{v}\=v_0-\langle v\rangle$ to the velocity field and $\Delta{a}\=2(v_0-\langle v\rangle)/\Delta{t}$ to its acceleration counterpart ($a\=\dot{v}$ and $\Delta{a}$ is the small shift in $a$) in computational nodes along the interface (with opposite signs for nodes belonging to the top/bottom body, to preserve the shear symmetry). This amounts to the application of a spatially-homogeneous external stress on the top/bottom nodes along the interface. This stress is typically very small, less than one percent of the initial frictional stress $\tau_0\=\sigma_0 f_{\rm ss}(v_{0})$, yet it nevertheless ensures that the average driving condition of Eq.~\eqref{eq:avr_v} is nearly satisfied at all times.

\section{The effective fracture energy}
\label{app:Gc}

The effective fracture energy $G_{\rm c}$ of interfaces obeying rate-and-state friction law is self-selected by the interfacial dynamics. In order to test the pulse equation of motion of Eq.~\eqref{eq:eom_prediction}, we computed $G_{\rm c}$ using two different methods. The first method relies on an approximated mapping of slip-weakening (rather than rate-weakening) friction laws --- characterized by a slip-weakening distance --- to rate-and-state ones, as discussed in~\cite{Bizzarri2003}.  In~\cite{Bizzarri2003} it has been shown that an effective slip-weakening distance $\delta^{\rm eff}_{\rm c}$ can be extracted in rate-and-state friction calculations, and that it approximately follows the relation $\delta^{\rm eff}_{\rm c}\!\approx\!D\ln(v_{\rm r}/v_{\rm bg})$,
where $v_{\rm bg}$ is the background slip velocity that a rupture mode propagates into and $v_{\rm r}$ is the residual slip velocity left behind it. This relation has been later used in~\cite{Rubin2005,Ampuero2008} to obtain an approximate expression for the effective fracture energy in the form $G_{\rm c}\!\approx\!\tfrac{1}{2}D b f_0\sigma_0\ln^2(v_{\rm r}/v_{\rm bg})$.
We used this estimate for $G_{\rm c}$, with $v_{\rm bg}\= v_{\rm s}$ and $v_{\rm r}\=v_0$, in Fig.~\ref{fig:fig5}. The results strongly supported the prediction in Eq.~\eqref{eq:eom_prediction}.

\begin{figure}[ht!]
\includegraphics[width=0.5\textwidth]{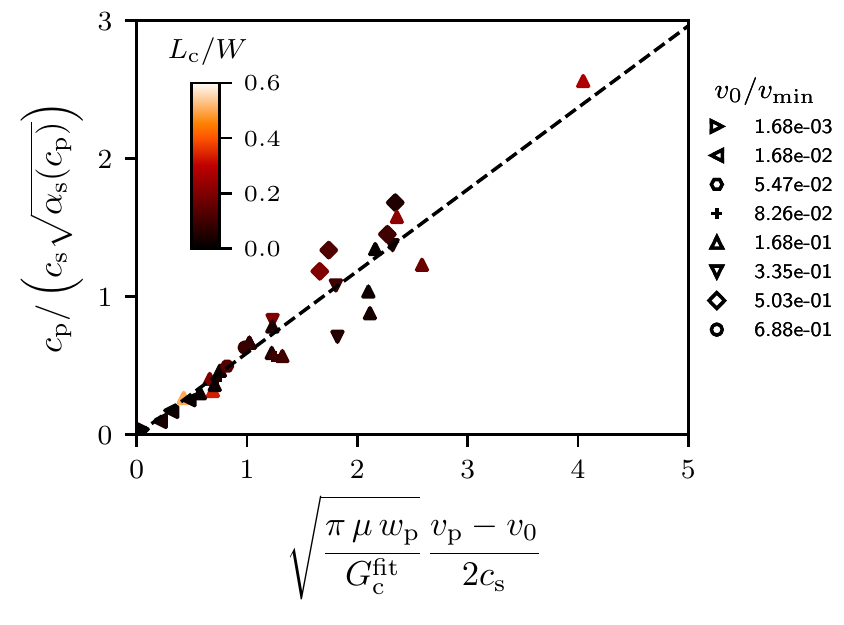}
\caption{Testing the theoretical prediction of Eq.~\eqref{eq:eom_prediction}, using the same data as in Fig.~\ref{fig:fig5}, but this time with $G_{\rm c}$ being estimated from the fit of the singular fields near the rupture edge, which is denoted by $G^{\rm fit}_{\rm c}$ (see text for discussion). The results reveal reasonably good agreement with the prediction, with an ${\cal O}(1)$ pre-factor, similarly to the results of Fig.~\ref{fig:fig5}.}
\label{fig:figSM3}
\end{figure}

The second method relies on the extraction of the nearly singular fields in the vicinity of the leading edge of the propagating pulse, shown in Fig.~\ref{fig:fig4} to be reasonably well approximated by the classical square root singular fields of LEFM. In particular, such a procedure allows to extract the stress intensity factor $K_{\rm III}$ by fitting simulational data to Eq.~\eqref{eq:sqrt}. Invoking then the leading edge energy balance of Eq.~\eqref{eq:EOM}, one can obtain an estimate of the effective fracture energy, which in this context is denoted by $G^{\rm fit}_{\rm c}$. In Fig.~\ref{fig:figSM3}, we present the very same data (and combinations of physical quantities in the $x$ and $y$ axes) as in Fig.~\ref{fig:fig5}, but this time using $G^{\rm fit}_{\rm c}$ for $G_{\rm c}$, estimated from fitting the singular fields near the pulse leading edge, as just described. The results yet again agree with the prediction in Eq.~\eqref{eq:eom_prediction} reasonably well, demonstrating the validity of the latter independently of the method used to estimate the effective fracture energy.

\section{The SW friction law}
\label{app:weakening}

Many studies available in the literature
employ a rate-and-state friction law that does not feature rate-strengthening at relatively high slip velocities (and hence no local minimum as observed in Fig.~\ref{fig:fig1}b)~\cite{Perrin1995,Zheng1998,Rubin2005,Ampuero2008,Nielsen2017}. Such a behavior is exhibited by the strengthening-weakening (SW) friction law discussed in Appendix~\ref{app:friction_laws} (which corresponds to using $f^{\rm SW}(|v|,\phi)$ in Eq.~\eqref{eq:n_shape}), whose steady-state curve is shown in Fig.~\ref{fig:figSM1} (dashed green curve). For completeness, we performed mode-III BIM calculations ($H\!\to\!\infty$) with the SW friction law. We find qualitatively similar results to those obtained for the N-shaped law, i.e.~coarsening dynamics towards a pulse train with periodicity set by the system length, $W_{\rm p}\=W$.  A representative movie of such calculations is available in~\cite{Movies}, see Appendix~\ref{app:movies} for details. We do note that the SW friction law leads to sharper pulses and typically features higher slip and propagation velocities compared to their N-shaped counterparts (not shown).\\

\section{Movies}
\label{app:movies}

Movies are available at~\cite{Movies} as follows:
\begin{itemize}
    \item M1.mp4: Infinite domain with the N-shaped friction law, see Fig.~\ref{fig:fig2}a
    \item M2.mp4: $H\!<\!L_{\rm c}<W$ with the WS friction law, corresponding to Fig.~\ref{fig:fig3}
    \item M3.mp4: $H\!<\!L_{\rm c}\!<\!W$ with the WS friction law, corresponding to Fig.~\ref{fig:fig3} in the co-moving frame of the pulse.
    \item M4.mp4: $H\!\simeq\!W\!\gg\!L_{\rm c}$ with the WS friction law.
    \item M5.mp4: Infinite domain with the SW friction law, see Appendix~\ref{app:weakening}.
\end{itemize}


%

\end{document}